\documentclass[aip,jcp,reprint]{revtex4-2}
\usepackage{graphicx} 
\usepackage{amsmath}
\usepackage{amssymb}
\usepackage[utf8]{inputenc}
\usepackage[T1]{fontenc}
\usepackage{mathptmx}
\usepackage{etoolbox}
\usepackage[main=english]{babel}

\makeatletter
\def\@email#1#2{%
 \endgroup
 \patchcmd{\titleblock@produce}
  {\frontmatter@RRAPformat}
  {\frontmatter@RRAPformat{\produce@RRAP{*#1\href{mailto:#2}{#2}}}\frontmatter@RRAPformat}
  {}{}
}%
\makeatother
\usepackage{tabularx}
\usepackage{array} 
\usepackage{booktabs}
\usepackage{hyperref}

\begin{document}
\title{Quantifying Local Point-Group-Symmetry Order in Complex Particle Systems}
\author{Domagoj Fijan} 
\affiliation{Department of Chemical Engineering, University of Michigan, Ann Arbor, MI, USA}
\author{Maria R. Ward Rashidi}
\affiliation{Department of Materials Science \& Engineering, University of Michigan, Ann Arbor, MI, USA}
\author{Jenna Bradley}
\affiliation{Department of Materials Science \& Engineering, University of Michigan, Ann Arbor, MI, USA}
\author{Sharon C. Glotzer}
\affiliation{Department of Chemical Engineering, University of Michigan, Ann Arbor, MI, USA}
\affiliation{Department of Materials Science \& Engineering, University of Michigan, Ann Arbor, MI, USA}
\affiliation{Biointerfaces Institute, University of Michigan, Ann Arbor, MI, USA}
\email{sglotzer@umich.edu}


\begin{abstract}
Crystals and other condensed phases are defined primarily by their inherent symmetries, which play a crucial role in dictating their structural properties. In crystallization studies, local order parameters (OPs) that describe bond orientational order are widely employed to investigate crystal formation. Despite their utility, these traditional metrics do not directly quantify symmetry, an important aspect for understanding the development of order during crystallization. To address this gap, we introduce a new set of OPs, called Point Group Order Parameters (PGOPs), designed to continuously quantify point group symmetry order. We demonstrate the strength and utility of PGOP in detecting order across different crystalline systems and compare its performance to commonly used bond-orientational order metrics. PGOP calculations for all non-infinite point groups are implemented in the open-source package SPATULA (Symmetry Pattern Analysis Toolkit for Understanding Local Arrangements), written in parallelized C++ with a Python interface. The code is publicly available on GitHub at \url{https://github.com/glotzerlab/spatula}.
\end{abstract}

\maketitle

\section{Introduction}

Crystals form via symmetry-breaking phase transitions, making the interrogation of local symmetry and how it develops along crystallization pathways essential for understanding and controlling the mechanisms by which a system transforms into a crystal.\cite{izyumov_phase_1990} 
Space group symmetry  -- the defining property of crystals -- is a combination of translational and point group (PG) symmetries, closely linked to the Wyckoff site symmetry of individual particles.\cite{souvignier_space_2016} 
Historically, order parameters (OPs) that quantify bond orientational order (BOO) have been used to study structure evolution during nucleation and growth.\cite{russo_crystal_2016,lee_entropic_2019} 
This approach ignores relative neighbor distances and partially decouples neighbor orientation order from local density.\cite{russo_crystal_2016} 
Common examples include Steinhardt Order Parameters (SOP)\cite{steinhardt_bond-orientational_1983} and their derivatives, such as Minkowski Structure Metrics (MSM)\cite{mickel_shortcomings_2013,lechner_accurate_2008} and the Lechner-Dellago approach,\cite{lechner_accurate_2008} as well as recent approaches for comparison of local BOO via normalized correlation between spherical harmonics.\cite{varela-rosales_solid-angle_2025}
For atomistic systems, the Smooth Overlap of Atomic Positions (SOAP) descriptors are often used to describe bond-orientational and radial correlations between atoms.\cite{bartok_representing_2013,caro_optimizing_2019,caruso_time_2023,darby_compressing_2022}
OPs such as these have also been used as descriptors in machine learning (ML) applications, providing insights into complex structural environments.\cite{bartok_machine_2017,grisafi_symmetry-adapted_2018,adorf_analysis_2020,chung_data-centric_2022,larsen_robust_2016,chen_collective_2018,chen_direct_2021,kashyap_perfect_2019} 
A complementary qualitative metric is the bond orientational order diagram (BOOD), which tracks the evolution of system-averaged BOO by mapping neighbor directions to the unit sphere and averaging over particles.\cite{damasceno_predictive_2012,russo_understanding_2014,schultz_symmetry_2015,haji-akbari_degenerate_2011,tracey_programming_2021}

All of these OPs share a common objective: to describe the evolution of (typically local) order associated with symmetry-breaking transitions.
However, all of these OPs trade one advantage for another.
For example, interpretability is strong in OPs such as BOODs, but they are qualitative at best. OPs based on local structure matching, such as Polyhedral Template Matching (PTM)\cite{larsen_robust_2016} or environmental matching,\cite{teich_identity_2019} are not bounded, and their values depend on both the system’s length scale and density, making comparisons across systems difficult.
Furthermore, OPs such as SOP, MSM and ones learned from ML are quantitative, but are often difficult to interpret and lack physical insight.

A local order parameter that directly quantifies PG symmetry would be invaluable for investigating crystallization pathways, providing a more direct measure of emergent symmetry than existing OPs, which serve as indirect proxies. 
Several approaches tailored to molecular systems have been proposed, often based on point-set registration\cite{alon_improved_2018,pinsky_continuous_1998,zabrodsky_continuous_1992}.
Since PG symmetry is of crucial importance for accelerating \textit{ab intio} and density functional theory (DFT) calculations in electronic structure theory, many approaches have been devised to determine the point groups of molecules and other atomic neighborhoods.\cite{leforestier_computer_1976,cao_molecular_1989,orlando_full_2014,largent_symmetrizer_2012,hanson_jmol_2010}
However, these methods share the drawbacks of template-matching and registration approaches.
Recent attempts to incorporate symmetry more explicitly include adapting SOPs to be symmetry-aware\cite{logan_symmetry-specific_2022} and defining PG symmetries of system-averaged BOODs.\cite{engel_point_2021,mbah_early-stage_2023}

Beyond crystals, symmetry also helps quantify mechanics in disordered solids, where local inversion symmetry predicts stability and broken symmetry drives nonaffine softening.\cite{liu_local_2022,zaccone_approximate_2011,zaccone_elasticity_2023}
Likewise, studies of liquid structure would benefit from symmetry-based order parameters when comparing to candidate crystal structures, for example in linking liquid structure to crystalline phase diagrams,\cite{fijan_liquid_2019} revealing crystal-like preordering and its impact on nucleation and growth,\cite{hu_revealing_2022, lee_entropic_2019} and connecting liquid heterogeneity to nucleation pathways across different materials systems.\cite{hallett_local_2018,bowles_influence_2022,rogal_controlling_2023,kurita_drastic_2019} 

In this work, we develop a general framework for constructing continuous, bounded, and interpretable OPs that quantify symmetry order as a non-binary value at a point of interest. The main idea of the proposed framework is to measure symmetry in a continuous fashion by replacing point particles with Gaussian functions, symmetrizing the environment of interest, and comparing the original and symmetrized environments through the overlap of Gaussian functions. By applying and averaging over symmetry operations in a given point group, we quantify the deviation of the environment from a perfectly symmetric configuration with that point-group symmetry. We implemented this methodology in the open-source, parallelized SPATULA (Symmetry Pattern Analysis Toolkit for Understanding Local Arrangements) toolkit written in C++ with a Python interface, which provides access to the broad scientific Python ecosystem as well as the speed of a compiled language. In our study, we benchmarked PGOP across simple (e.g., FCC, HCP) and complex (e.g., A15, $\mathrm{\gamma}$-brass, pyrochlore) crystal systems, evaluating its sensitivity to noise and its ability to successfully distinguish different Wyckoff-site environments, and comparing its performance to MSM in detecting order.  Additionally, we applied PGOP to study nucleation of the well-known Lennard-Jones system using molecular dynamics. SPATULA can be used to calculate PGOPs for all non-infinite point groups and is publicly available on GitHub.

\section{Theory and Methods}

\subsection{Theoretical Development}

\subsubsection{Formulation of a continuous symmetry order parameter}

\begin{figure*}
    \centering
    \includegraphics[width=\textwidth]{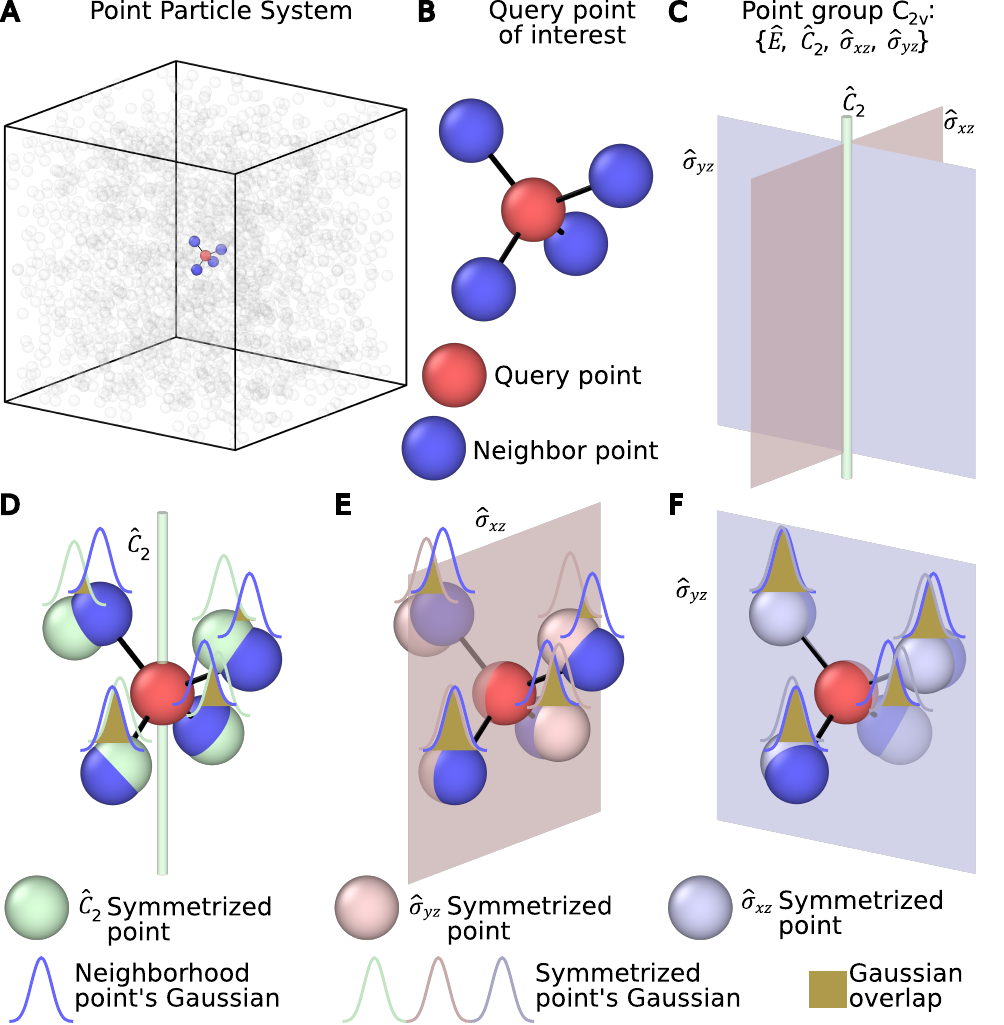}
    \caption{
    Steps for computing the point group order parameter (PGOP) for a given query point (red sphere).
    (\textbf{A}) The system of semi-transparent white points with the query point highlighted in red.
    (\textbf{B}) The nearest neighbors of the query point (blue) are identified.
    (\textbf{C}) The chosen point group, $\mathrm{C_{2v}}$, with its symmetry elements: the two-fold rotation axis $\hat{C}2$ (green) and the mirror planes $\hat{\sigma}_{xz}$ (light red) and $\hat{\sigma}_{yz}$ (light blue).
    (\textbf{D–F}) The initial neighborhood configuration (blue) is symmetrized with respect to each symmetry element, yielding symmetrized neighborhoods colored according to the generating operator: $\hat{C}2$ (green, \textbf{D}), $\hat{\sigma}_{xz}$ (light red, \textbf{E}), and $\hat{\sigma}_{yz}$ (light blue, \textbf{F}).
    The 1D Gaussian cartoons represent the 3D Gaussians actually used in the calculation. Overlaps between each symmetrized neighborhood and the original neighborhood are shown in tan. The PGOP value is obtained as the average sum of these overlaps.
    }
    \label{fig:method}
\end{figure*}

To distinguish between symmetric and asymmetric arrangements of points, we utilized  symmetrization following work by Engel.~\cite{engel_point_2021}
Symmetrization can be applied on a per (symmetry) operator basis or a per group basis (group action).
It involves applying an operator in a given representation to particle positions in the same representation.
This process leaves symmetric configurations unchanged.
In other words, if the starting coordinates are transformed into new coordinates that align with the starting coordinates, the arrangement is considered symmetric with respect to the symmetry operation applied.
If points do not perfectly align, the starting coordinates were not in a symmetric arrangement. 
However, the combined set of original and transformed points will now exhibit symmetry with respect to the applied operation, hence the name ``symmetrization''.
In $\mathbb{R}^3$, group action is applied by multiplying the particle coordinates with a matrix representation of each symmetry operation.

Having applied the group action to an arrangement of points, we can assess the degree of symmetry order of the arrangement by computing the overlap between the symmetrized and original point sets. This overlap quantifies how closely the initial configuration approximates perfect symmetry. When particles are represented as delta functions as in work by Engel,\cite{engel_point_2021} the overlap yields a value of 1 for a symmetric configuration and 0 for an asymmetric one. While this discreteness is fine for a global system-averaged OP that accumulates contributions from many points,\cite{engel_point_2021} the discrete nature of the delta function is insufficient for a single-site, local per-particle OP.
To convert from a discrete to a continuous measure, we replace the delta functions representing particle locations with a smooth continuous distribution.
This approach was first proposed in work by Butler,\cite{butler_development_2024} where Fisher functions were used to represent points, which are then projected to a unit sphere centered on a particle to construct a local BOOD.
In this work, rather than restricting the representation to the unit sphere (where spherical harmonics provide a convenient basis), we work directly in three-dimensional Euclidean space so that radial information is retained.
We replace Dirac delta functions by isotropic Gaussian densities, for which the required overlap integrals can be evaluated analytically.
Working in the embedding space avoids the explicit complex-valued basis set expansion used, e.g., in Butler\cite{butler_development_2024} and Engel\cite{engel_point_2021}) and reduces overhead by enabling single-precision implementations for improved performance.
We then compare the original and symmetrized Gaussian densities using a distribution similarity measure, rather than inner-product-based normalized correlations used in those spherical-harmonic-based formulations.\cite{butler_development_2024,engel_point_2021}
For normalized Gaussian densities, this yields closed-form expressions and keeps the resulting order parameter bounded in $[0,1]$.

In three dimensions, most point groups are defined only up to an overall rotation, so the orientation of the group (in particular its principal axis) is a free parameter.
Consequently, symmetrization-based symmetry measures, including our approach and related BOOD-based formulations,\cite{butler_development_2024,engel_point_2021,mbah_early-stage_2023} require an optimization over $R\in SO(3)$ to find the group orientation that maximizes the similarity between the original and symmetrized representations.
Similar $SO(3)$ alignment steps also appear in order parameters that match local environments to templates, including Polyhedral Template Matching (PTM)\cite{larsen_robust_2016}, environmental motif matching,\cite{teich_identity_2019} cluster-alignment based motif scores,\cite{fang_atomistic_2010,sun_crystal_2016} and the local-order metric (LOM).\cite{martelli_local-order_2018}
In practice, these optimizations are handled via best-fit rigid rotations inside an assignment search, coarse-to-fine orientation sampling with local refinement, or accelerated global scans based on FFT and fast $SO(3)$ Fourier transforms.\cite{griffiths_optimal_2017}
Here, we use a coarse grid search followed by steepest-descent refinement, following Butler.\cite{butler_development_2024}

\subsubsection{Point group order parameter implementation}

PGOP calculations require specifying the target point group symmetry, Gaussian widths, and the positions of neighboring particles (neighbor list) with respect to the point of interest (query point; see Figure \ref{fig:method} \textbf{A} and \textbf{B}). 
Neighborhoods may also be evaluated at arbitrary (non-particle) locations.
Symmetry operations are computed in $\mathbb{R}^3$ representation for the selected point group.
A detailed summary of the symmetry operations for each point group, along with their matrix representations, is provided in the Supplementary Material SM1.

Neighborhoods are constructed using \textit{freud}\cite{dice_analyzing_2019,ramasubramani_freud_2020} for each query point and then symmetrized via group action. 
For each symmetry operator in the selected point group (Figure \ref{fig:method} \textbf{C}), we apply its $\mathbb{R}^3$ representation to the computed neighborhood coordinates relative to the query point, generating the corresponding symmetrized neighborhood (Figure \ref{fig:method} \textbf{D}-\textbf{F}).

Next, for each symmetrized configuration generated by its corresponding symmetry operator, we place normalized Gaussians at the particle positions in both the symmetrized and original neighborhoods (Figure \ref{fig:method} \textbf{D}-\textbf{F}).
To quantify similarity between symmetrized and original neighbor positions, we use the Bhattacharyya coefficient (BC).\cite{bhattacharyya_measure_1943,bhattacharyya_measure_1946,kashyap_perfect_2019}
BC measures the similarity between two distributions and is closely related to their overlap.
Derivations for the BCs involving Gaussian distributions are provided in the Supplementary Material SM2.
For each symmetrized point, we compute BCs to all original positions and retain the maximum value only.
This is required in order to keep the OP bounded in the [0,1] range.
The final PGOP value is the average of these maxima over all symmetrized particles and symmetry operators.

We then optimize the orientation of the principal axis of symmetry for the specified point group to maximize the PGOP value.
This is done via $SO(3)$ optimization.
The procedure used in this work employs the same approach discussed in thesis by Butler.\cite{butler_development_2024}

The above steps can be summarized in the following expression for the PGOP:
\begin{widetext}
    
\begin{equation}
   \mathrm{PGOP} (r,R,S,Q) = \frac{1}{N_R N_S} \max_Q \Bigg[\sum_S \sum_{R_i} \max_{R_j} \bigg[\mathrm{BC}\Big\{F(R_j-r),F\big(Q_iSQ_i^T(R_i-r)\big)\Big\}\bigg]\Bigg]
\end{equation}
\end{widetext}
where $r$ is the location of the query point in space, $R$ is a set of $N_R$ neighbors around the query point, $S$ is the set of $N_S$ symmetry operators defined by a PG symmetry, $Q$ is the set of rotation matrices used in the optimization, $R_i$ and $R_j$ are single neighbor positions, BC is the Bhattacharya coefficient, $F$ is a function (e.g. a Gaussian) centered at the coordinates of points, and $Q_i$ is a single rotation matrix.

\subsubsection{Computational complexity and performance}

The total computational complexity  scales as $O(K \times G \times M^2 \times N)$ for a system containing $N$ points, with $M$ neighbors in a point's neighborhood, $G$ symmetry elements in the point group, and $K$ optimization steps. In contrast, classical order parameters like the $\mathrm{MSM}_l$ scale as $O(l \times M \times N)$ for an expansion in spherical harmonics truncated at $l$. While $l$ is typically a fixed integer from 1 to 12, the group order $G$ varies  from 2 to as many as 120 for the icosahedral group: however, the largest crystallographically compatible point group, $\mathrm{O_h}$, is the typical upper bound with order 48. Supplemental Figure S3 shows benchmarks highlighting the impact of group size on order parameter performance. Because PGOP operates on Cartesian positions rather than spherical harmonics, we do not require an expansion in $l$ and are able to use single precision coordinates rather than complex double precision needed for spherical harmonics. Our benchmarks indicate that use of single precision provides a significant boost to PGOP's performance without measurably impacting the accuracy of the results. Additional, we are able to efficiently employ Single Instruction Multiple Data (SIMD) parallelism, which is discussed further in the Supplementary Material SM5.

The performance of PGOP was compared against the MSM with spherical harmonic $l=6$ ($\mathrm{MSM}_6$) for a system of 13,824 particles and a Voronoi neighbor list. \textit{freud}'s implementation of $\mathrm{MSM}_6$ is capable of processing $(2.664 \pm 0.022) \cdot10^5 $ particles per second on a single node of the Purdue Anvil supercomputer (128 cores). Using the same computational resource, PGOP is able to quantify the degree of octahedral point group $\mathrm{O_{h}}$ symmetry for $(0.261 \pm 0.006)\cdot10^5$ particles per second: within a factor of ten of $MSM_6$ despite extracting significantly more information. On the consumer Apple M1 Pro processor with 8 cores, we are able to quantify the degree of $\mathrm{D_{3h}}$ order for 2,080 frames in a system of 13,824 particles per hour. This is a factor of 23 times slower than $\mathrm{MSM}_6$, but is sufficient to characterize large volumes of data on modest resources. All calculations included 600 hyperspherical grid optimization points combined with gradient descent.  We note that, in the case where no numerical optimization is required (for example when the orientation of a structure is known \textit{a priori} or a particular measurement orientation is desired), PGOP's performance is within $50\%$ of $\mathrm{MSM}_6$ and is faster than $\mathrm{MSM}_{12}$ for all crystallographic point groups. Full strong scaling results are provided with Supplementary Figure S4.

\subsubsection{Extensions}
The above framework can be applied not only to PG symmetries but also to individual symmetry operators (e.g., rotation, reflection, translation) and any sets formed from these operators, including space groups.
This generality arises because the approach is independent of the symmetry operation type, provided it can be represented in $\mathbb{R}^3$.
The generalization to any number of spatial dimension only requires a way to represent the desired symmetry operators in $\mathbb{R}^N$.
The framework can also be used to determine the PG symmetry order associated with other quantities defined at individual points.

\subsubsection{Extension to local Bond Orientational Order Diagrams}

To compute the point-group symmetry order of a local (single particle environment) BOOD, we first construct the original neighborhood in $\mathbb{R}^3$, apply the point-group symmetrization to obtain a symmetrized neighborhood, and then project both neighborhoods onto the unit sphere.
This procedure differs from Butler\cite{butler_development_2024} and Engel,\cite{engel_point_2021} where the neighborhood is first projected to the sphere to form a BOOD (local in Butler, global in Engel), and symmetrization is then carried out in the spherical-harmonic representation.\cite{butler_development_2024,engel_point_2021}
After projection, the discrete points on the sphere are replaced by Fisher distributions following Butler,\cite{butler_change_2024} and the similarity between the original and symmetrized spherical densities is computed using the BC, rather than inner-product-based correlation measures used, e.g., in Butler\cite{butler_development_2024} and Engel.\cite{engel_point_2021}
The PGOP-BOOD, as well as approaches by Butler\cite{butler_development_2024} and Engel,\cite{engel_point_2021} effectively eliminates distance information from the PG symmetry order calculation.
Derivations for the BCs involving Fisher distributions are provided in the Supplementary Material SM2.

\subsection{Computational details}

\subsubsection{Application of order parameters on constructed crystal systems with noise}

To quantify the performance of the PGOP, we set out to apply it on several crystal systems.
All crystal supercells were generated using the \textit{freud}'s data module.\cite{ramasubramani_freud_2020,dice_analyzing_2019}
Coordinates for hexagonal close-packed (HCP) and complex crystals were sourced from AFLOW\cite{mehl_aflow_2017,hicks_aflow_2019,hicks_aflow_2021,eckert_aflow_2024}. 
Table \ref{tab:crystals} summarizes the prototypes and Wyckoff site symmetries in constructed complex crystals and the HCP crystal.

To construct the supercells, the unit cells were replicated in each dimension for all crystals for a given number of repeat counts.
For simple crystals, super cells were constructed with 6 repeats for face centered cubic (FCC) crystal, 8 for body centered cubic (BCC) and HCP crystals, and 10 for simple cubic (SC) crystal.
For complex crystals, repeat counts were 5 for A15, 4 for $\gamma$-brass, and 3 for pyrochlore.
The pyrochlore crystal was generated with a primitive cell volume of 2 (smaller compared to Aflow prototype).
Noise was introduced by adding random displacements to particle positions, drawn from a Gaussian distribution ($\Sigma = 0.02$), while ideal crystals were constructed with no noise ($\Sigma = 0$).
An ideal gas configuration served as a reference and was constructed using \textit{freud}'s data module.

\begin{table*}[ht]
\centering
\caption{Crystal prototypes used for construction of crystal supercells}
\label{tab:crystals}
\begin{tabularx}{\textwidth}{
    >{\raggedright\arraybackslash}p{0.1425\textwidth}  
    >{\raggedright\arraybackslash}p{0.45\textwidth}  
    >{\raggedright\arraybackslash}X                  
}
\hline
Crystal & AFLOW prototype & Wyckoff sites \\
\midrule
HCP & A\_hP2\_194\_c-001 & 2c (D$_{\mathrm{3h}}$)  \\
\midrule
A15 & A3B\_cP8\_223\_c\_a‑001 & 2a (T$_\mathrm{h}$), 6c (D$_{\mathrm{2d}}$) \\
\hline
$\gamma$‑brass & A5B8\_cI52\_217\_ce\_cg‑001 & 8c (C$_\mathrm{3v}$), 12e (C$_\mathrm{2v}$), 24g (C$_\mathrm{s}$) \\
\hline
pyrochlore & A2BCD3E6\_cF208\_203\_e\_c\_d\_f\_g‑001 & 16c (S$_\mathrm{6}$), 16d (S$_\mathrm{6}$), 32e (C$_\mathrm{3}$), 48f (C$_\mathrm{2}$), 96g (C$_\mathrm{1}$)\\
\hline
\end{tabularx}
\end{table*}

To compare the sensitivity of PGOP, PGOP-BOOD, and MSM($l=6$) to noise, we computed these order parameters for HCP and FCC crystals perturbed by increasing levels of Gaussian noise with standard deviation $\Sigma$. This approach allows us to probe the sensitivity of each OP to residual order in regions of extreme disorder, where a simulated system would otherwise undergo a phase transition. Gaussian displacement noise is physically motivated as a coarse-grained model for colloidal systems in the overdamped Brownian diffusion limit, where displacements are Gaussian,\cite{bian111YearsBrownian2016} and is also consistent with the weak-collision regime, where Gaussian statistics follow from the central limit theorem.\cite{dubkovNonlinearBrownianMotion2009}
MSM are Voronoi‐based, rotation‐invariant order parameters that weight each neighbor’s bond orientational order contribution by the relative area of the shared Voronoi facet, yielding a continuous, parameter-free generalization of SOPs.
For all order parameters computed for sensitivity comparison, the neighbor list was constructed using Voronoi tessellation, as implemented in the \textit{freud} package.
PGOP and PGOP-BOOD were computed using point group O$_\mathrm{h}$ for FCC and D$_\mathrm{3h}$ for HCP.
For PGOP, a Gaussian width of $\sigma = 0.075$ was used, while a $\kappa$ value of 50 was used for PGOP-BOOD.

To distinguish particle environments in different simple crystals by analyzing the structure of their local neighborhoods, we generated SC, BCC, FCC, and HCP crystals both without noise and with Gaussian noise ($\Sigma = 0.02$).
PGOP was computed for point groups O$_\mathrm{h}$ and D$_\mathrm{3h}$.
All PGOP calculations used a Gaussian width of $\sigma = 0.075$, and neighbor lists were constructed using the Relative Angular Distance (RAD) method,\cite{higham_locally_2016,higham_overcoming_2018} as implemented in \textit{freud}.\cite{ramasubramani_freud_2020,dice_analyzing_2019}

To distinguish particles at different Wyckoff sites in A15 and $\gamma$-brass crystals, we analyzed crystal configurations without noise and with Gaussian noise ($\Sigma = 0.02$).
PGOP was computed for point groups T$_\mathrm{h}$ and D$_\mathrm{2d}$ in A15, and for point groups C$_\mathrm{2v}$ and C$_\mathrm{3v}$ in $\gamma$-brass.
All PGOP calculations used a Gaussian width of $\sigma = 0.075$, and neighbor lists were constructed using the Relative Angular Distance (RAD) method,\cite{higham_locally_2016,higham_overcoming_2018} as implemented in \textit{freud}.\cite{ramasubramani_freud_2020,dice_analyzing_2019}

To study the impact of the initial neighborhood definition on PGOP results, we computed PGOP values using four different neighbor lists for the pyrochlore crystal.
PGOP was calculated for point groups $\mathrm{C_2}$ and $\mathrm{C_3}$ using $\sigma = 0.075$.
Neighbor lists were constructed in \textit{freud} using Solid Angle Nearest Neighbors (SANN),\cite{van_meel_parameter-free_2012} Radial Angular Distance (RAD),\cite{higham_locally_2016, higham_overcoming_2018} Voronoi,\cite{voronoi_nouvelles_1908} and a standard ball query with a radial cutoff of 1.1 computed via an axis-aligned bounding box (AABB) query.\cite{howard_efficient_2016}
SANN, RAD, and Voronoi are parameter-free neighbor lists that define the first coordination shell geometrically and can be used to study short-range order.
In contrast, standard neighbor lists rely on user-defined parameters to set the length scale at which neighbors are identified, enabling the definition of larger neighborhoods — typically using a radial cutoff or a pre-defined number of nearest neighbors.

\subsubsection{Simulations of crystals with thermal noise}

Thermal noise calculations for FCC and HCP crystals were performed with HOOMD-blue\cite{anderson_hoomd-blue_2020} version 4.7 in a signac-managed workflow\cite{adorfSimpleDataWorkflow2018} using Lennard-Jones potentials. Two sets of simulations were run: FCC set at $\rho=1.00$ with potential set ($r_c=1.8$, shifted truncation, no tail correction) and HCP set at $\rho=1.10$ with($r_c=2.0$, shifted truncation, no tail correction). Temperatures were explored over the range $kT \in [0, 3]$ with a step size of $0.1$. Initial particle configurations were generated from ideal FCC or HCP unit cells, replicated to approximately $N\sim 10^3$ particles (exact $N$ from integer replication at fixed density), and integrated in the NVT ensemble with a Bussi thermostat\cite{bussi_canonical_2007} and $\Delta t=0.002$. All values reported are in reduced units. Each run consisted of $50,000$ equilibration and $200,000$ production steps, with frames saved every 1000 steps. For each saved frame, neighbors were defined by a Voronoi construction and three order metrics were computed: MSM$_6$, PGOP ($\sigma_{\mathrm{PGOP}}=0.075$), and PGOP-BOOD ($\kappa=50$). For FCC and HCP crystals we used $\mathrm{O_h}$ and $\mathrm{D_{3h}}$ groups, respectively, and particle-level values were pooled across frames to compute the means, standard deviations, and the range of values between the first and last quartiles (interquartile range, IQR). To quantify the loss of order due to thermal noise, we used fluid configurations generated at the same density and potential and near melting temperatures as baselines.

\subsubsection{Simulations of nucleation in Lennard-Jones system}

To study nucleation from a metastable LJ liquid, we performed molecular dynamics (MD) simulations in NVT ensemble using the Bussi thermostat\cite{bussi_canonical_2007} as implemented in HOOMD-blue version 4.7.0.\cite{anderson_hoomd-blue_2020}
The starting system configuration of 1000 particles in a simple cubic configuration of box length 13 was melted at temperature $T_\mathrm{melt}=10.5$ reduced units for 10,000 steps.
The system is then cooled to temperature $T=1.15$ and compressed to density $\rho=1.0$ (both in reduced units) over 35,000 steps.
Next, we equilibrate the system at target self-assembly conditions of $T=0.8$ and $\rho=1.0$ (both in reduced units) for 50,000 time steps.
After equilibration, we start the production run for 10,000,000 steps.
During the production run we run on-the fly detection of rare events using \textit{dupin}.\cite{butler_change_2024}
A specialized HOOMD custom action for capturing rare nucleation event subtrajectories at high temporal resolution is employed.
The RAD neighbor list and $\sigma=0.075$ was used for the PGOP calculations.

To obtain a highly time-resolved subtrajectory of a nucleation event, we employ a custom HOOMD action and a custom HOOMD burst writer. During the production MD run, we store the most recent 10,000 trajectory frames spaced 10 time steps apart in memory using a burst buffer. As the simulation progresses, the old frames are discarded and new frames are added to the burst buffer queue. For the duration of the simulation, every 10,000 steps we attempt to run \textit{dupin}'s online detection algorithm.\cite{butler_change_2024}
\textit{dupin}'s detection algorithm uses a one-dimensional signal constructed from the 50th lowest value of potential energy in the simulation system. This signal has a maximal signal length of 500. Detection is accomplished using the sweep detector with 3 maximal change points, a tolerance of 0.01 and a piecewise linear fitting using an L1 cost function. Once an event is detected, the burst buffer is dumped onto the disk, and the burst buffer resets. This approach allows us to create highly time-resolved subtrajectories for rare events of interest while avoiding having to save long subtrajectories prior to the onset of nucleation (see Supplementary Material Figure S1).

\section{Results}
\subsection{Sensitivity to order and noise}

\begin{figure*}
    \centering
    \includegraphics[width=\textwidth]{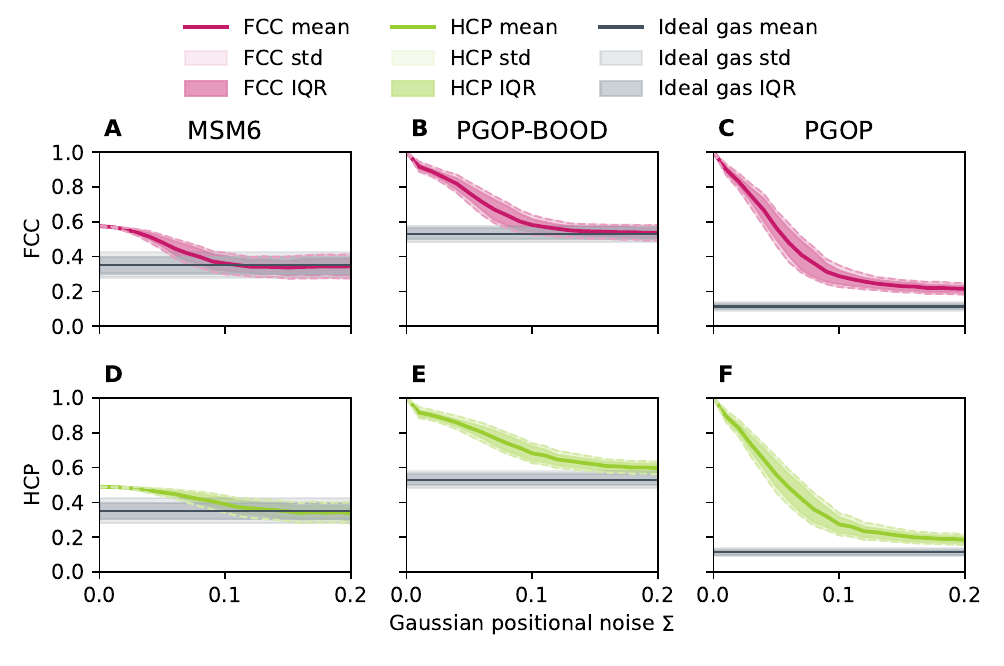}
    \caption{Order measured by MSM ($l=6$), PGOP, and PGOP-BOOD for FCC and HCP crystals with increasing positional noise. Line and shaded-region colors indicate FCC (magenta), HCP (yellow-green), and ideal gas (black/gray). For each dataset, the darker shaded region shows the spread between the first and last quartiles, and the lighter shaded region alongside dashed line shows the spread associated with one standard deviation. Panels \textbf{A} and \textbf{D} show MSM$_6$ for FCC and HCP, respectively. For FCC (top row), PGOP-BOOD (\textbf{B}) and PGOP (\textbf{C}) are computed for the point group $\mathrm{O_h}$. For HCP (bottom row), PGOP-BOOD (\textbf{E}) and PGOP (\textbf{F}) are computed for the point group $\mathrm{D_{3h}}$.}
    \label{fig:pgop_minkowski_noise_dependence}
\end{figure*}

We compare the sensitivity of PGOP, PGOP-BOOD, and MSM$_6$ in distinguishing noisy crystalline environments from an ideal gas (Figure \ref{fig:pgop_minkowski_noise_dependence}). Calculations were performed for FCC and HCP crystals under increasing positional noise ($\Sigma$) sampled from a normal distribution, using point groups $\mathrm{O_h}$ for FCC and $\mathrm{D_{3h}}$ for HCP.

The range of order parameter values between perfect crystal ($\Sigma=0$) and ideal gas is narrowest for MSM and widest for PGOP. For FCC, the MSM mean crosses the ideal gas mean at $\Sigma \approx 0.1$ (Figure \ref{fig:pgop_minkowski_noise_dependence} A), and at $\Sigma \approx 0.13$ for HCP (Figure \ref{fig:pgop_minkowski_noise_dependence} D). PGOP-BOOD is more robust: the crossing occurs at $\Sigma \approx 0.2$ for FCC (Figure \ref{fig:pgop_minkowski_noise_dependence} B), and does not occur within the sampled noise range for HCP (Figure \ref{fig:pgop_minkowski_noise_dependence} E), though it is expected near that value. PGOP outperforms both, maintaining higher contrast across the full noise range.

For FCC, PGOP’s ideal gas mean remains well below the lowest quantile at $\Sigma=0.2$ (Figure \ref{fig:pgop_minkowski_noise_dependence} C), indicating strong discrimination. For HCP, PGOP performs slightly worse (Figure \ref{fig:pgop_minkowski_noise_dependence} \textbf{F}), but still better than MSM and PGOP-BOOD. These results show that PGOP is significantly more sensitive to structural order than MSM and more sensitive than PGOP-BOOD, as expected given the latter’s insensitivity to neighbor distances.

In addition, we test the impact of thermal noise on the loss of order in equilibrium FCC and HCP crystals assembled from LJ systems (Supplementary Material Figure S5). The trends are consistent with the Gaussian noise results. PGOP and PGOP-BOOD remain discriminative between crystal and liquid structure up to the melting temperature, while MSM exhibits substantial overlap with the liquid distribution at lower temperatures for both FCC and HCP crystals. A detailed discussion is provided in Supplementary Material SM6.

\subsection{Discerning simple crystal environments using PGOP}

\begin{figure*}
    \centering
    \includegraphics[width=\textwidth]{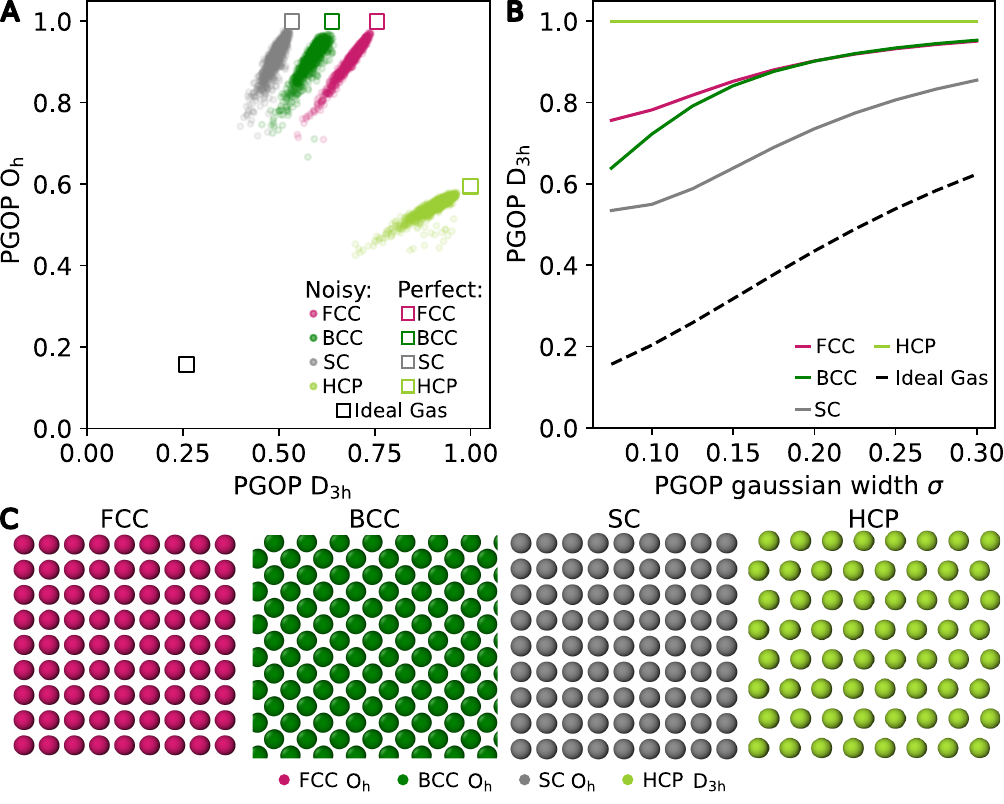}
    \caption{The application of PGOP for simple crystal identification. Each color corresponds to a specific neighborhood, and remains consistent across this figure: FCC in magenta, BCC in green, SC in gray, and HCP in yellow-green. \textbf{A}: Distribution of PGOP values (semi transparent points) for point groups $\mathrm{D_{3h}}$ and $\mathrm{O_h}$ for 4 different simple crystals (FCC, BCC, SC, HCP) constructed with Gaussian noise. The values for perfect crystals are given by square marks, while the ideal gas is denoted by a black square mark. \textbf{B} The change in PGOP $\mathrm{D_{3h}}$ value for perfect crystals as a function of Gaussian width ($\sigma$). Perfect crystal values are given with full lines, while ideal gas is denoted by dashed black line. \textbf{C}: The perfect simple crystals SC, FCC, BCC, HCP with particles colored by their environment. Every particle has a perfect value of symmetry for their respective crystals: perfect $\mathrm{D_{3h}}$ symmetry for HCP and perfect $\mathrm{O_h}$ symmetry for FCC, BCC and SC.}
    \label{fig:simple_crystals}
\end{figure*}

To showcase the utility of PGOP, we first explore its ability to distinguish different coordination environments in simple crystals (i.e., crystals with only one Wyckoff site). As examples, we apply PGOP to face centered cubic (FCC), body centered cubic (BCC), simple cubic (SC), and hexagonal close packed crystals (HCP) (perfect crystal structures are shown in Figure \ref{fig:simple_crystals} \textbf{C}).
We calculated the PGOPs for $\mathrm{D_{3h}}$ and $\mathrm{O_h}$ symmetry.

Despite the fact that all three cubic lattice crystals (FCC, BCC and SC) have the same octahedral PG symmetry of $\mathrm{O_h}$, with only HCP having a different PG symmetry of $\mathrm{D_{3h}}$, we can clearly see that just the combination of these two PGOP values comfortably separates all four crystal environments. This remains true even when noise is introduced into the system (Figure \ref{fig:simple_crystals} \textbf{A}).

\subsection{Discerning complex crystal environments using PGOP}

\begin{figure*}
    \centering
    \includegraphics[width=\textwidth]{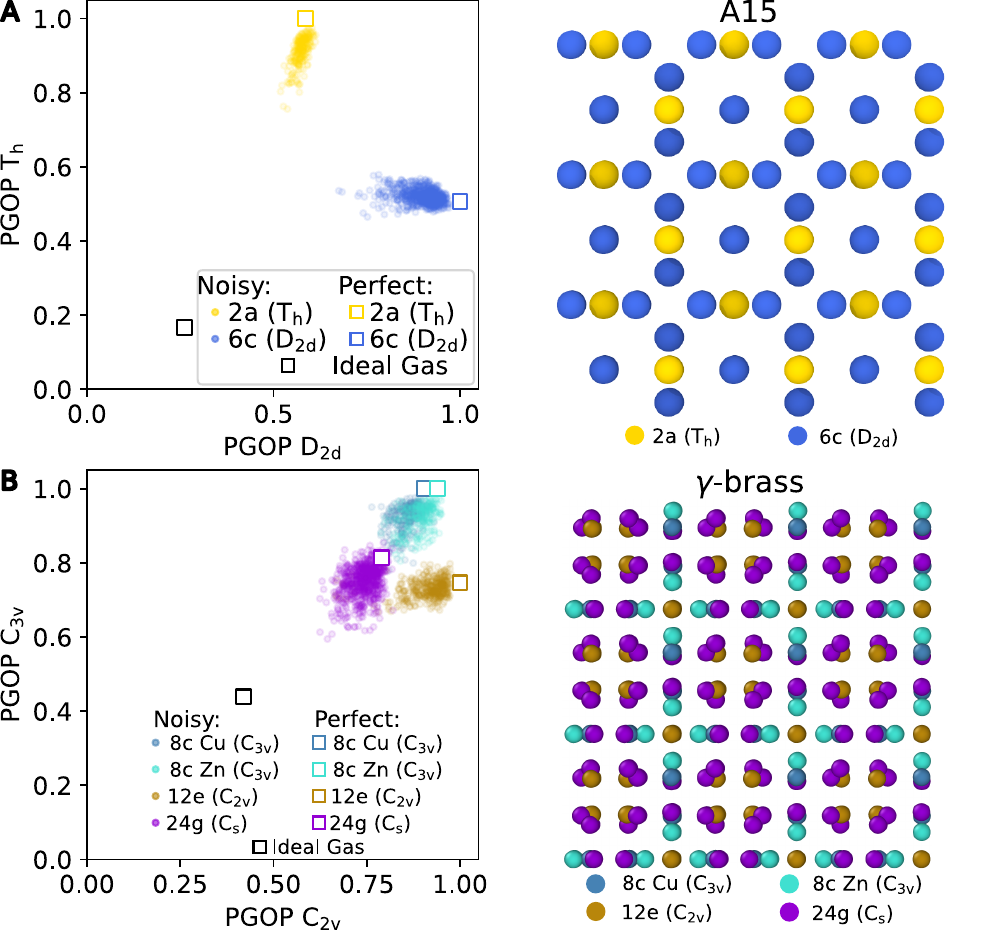}
    \caption{Using PGOP to identify local environments found in complex crystals. The particles within a noisy (\textbf{A}) A15 crystal and a noisy (\textbf{B}) $\mathrm{\gamma}$-brass crystal plotted by PGOP for two point groups found within the specified crystal ($\mathrm{T_h}$ and $\mathrm{D_{2d}}$ for A15; $\mathrm{C_{3v}}$ and $\mathrm{C_{2v}}$ for $\mathrm{\gamma}$-brass). PGOP plots are paired with snapshots of the perfect crystal colored by its Wyckoff site designation. These colors are consistent between the PGOP plots and the perfect crystal snapshot. For both crystals, the values for the perfect crystal are represented by squares, while circles represent values from the noisy crystals. The ideal gas value is represented by a black square.}
    \label{fig:complex_crystals}
\end{figure*}

Next, we attempt to distinguish different local environments in more complex crystals that possess multiple Wyckoff sites.
In Figure \ref{fig:complex_crystals}, we consider two complex crystal examples: the A15 crystal, which has 8 particles in its unit cell, and $\gamma$-brass, with a 52-particle unit cell. 
In the case of A15, we expect neighborhoods to have either $\mathrm{T_h}$ or $\mathrm{D_{2d}}$ symmetry based on their Wyckoff site designation (2a for $\mathrm{T_h}$ and 6c for $\mathrm{D_{2d}}$).
Using PGOP, we were able to completely separate the particles into their corresponding environments, even with noise.
For $\mathrm{\gamma}$-brass, the particles fall into one of three possible symmetries, $\mathrm{C_{3v}}$ (Wyckoff sites 8c), $\mathrm{C_{2v}}$ (Wyckoff site 12e), or $\mathrm{C_s}$ (Wyckoff site 24g).
The particles with $\mathrm{C_{3v}}$ symmetry can be further separated into two environments, based on the difference in the element identity in which they fall.
We chose to compute PGOP for only the point groups $\mathrm{C_{3v}}$ and $\mathrm{C_{2v}}$ since the point group $\mathrm{C_{s}}$ would not produce useful results as it is a subgroup of both $\mathrm{C_{3v}}$ and $\mathrm{C_{2v}}$.
When we use PGOP to distinguish these environments, we observe good separation between the environments exhibiting different symmetries, with partial overlap between environments that share the same symmetry (8c Zn and 8c Cu).

\subsection{Impact of distribution width on PGOP}

Two key parameters influence PGOP performance: the neighbor list and the distribution width, controlled by $\sigma$ (Gaussian) in PGOP or $\kappa$ (Fisher) in PGOP-BOOD. Figure \ref{fig:simple_crystals} \textbf{B} shows how PGOP values vary with $\sigma$ across several perfect crystal environments.

As $\sigma$ increases (or $\kappa \to 0$), the overlap between distributions increases, and PGOP values for all the crystals and ideal gas approach unity. This reduces PGOP's ability to distinguish local environments, as their values converge. Conversely, at low or intermediate $\sigma$, PGOP better resolves local environments, maximizing the separation between their values, while ideal gas values remain low.

In the limit $\sigma \to 0$ (or $\kappa \to \infty$), the distributions approximate delta functions. Overlap vanishes except for perfectly symmetric configurations, recovering the binary nature of symmetry. For non-symmetric configurations, the order parameter value in the limit $\sigma \to 0$ is never exactly zero, as the optimizer typically aligns at least one symmetrized particle with one original neighbor, yielding a small but finite value that depends on the point group and local geometry. In this regime, PGOP loses discriminatory power, as nearly all imperfect environments yield values near zero.

\subsection{Impact of the neighbor list choice on PGOP}

\begin{figure*}
    \centering
    \includegraphics[width=\textwidth]{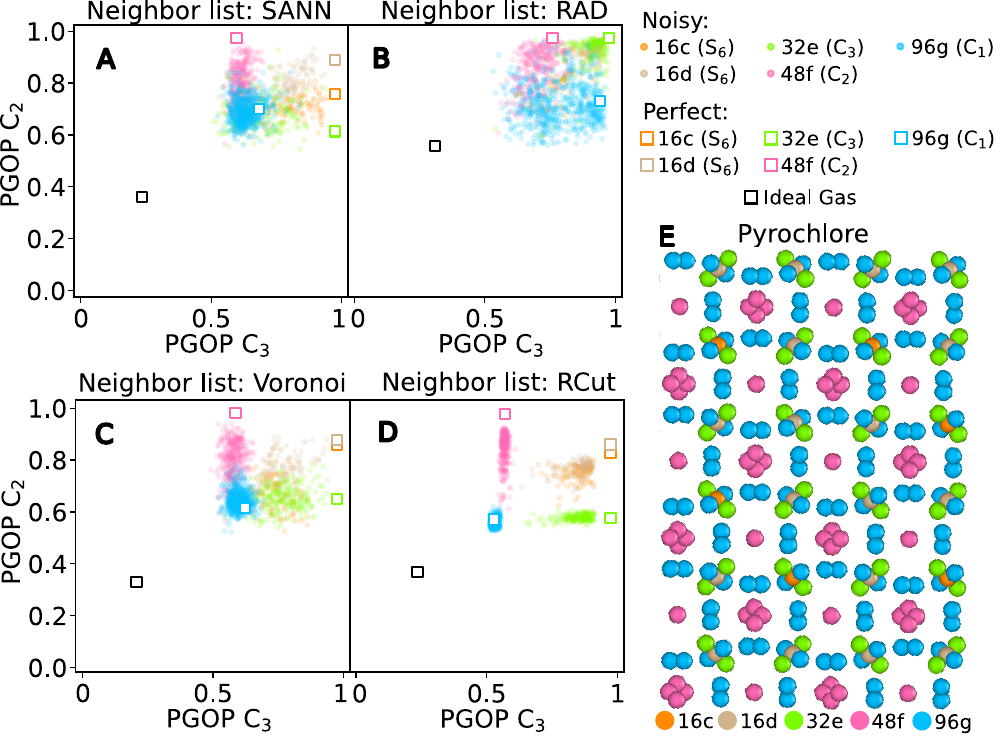}
    \caption{Impact of the neighbor list choice on PGOP results for pyrochlore crystal. Four panels display the particle values of PGOP for different neighbor lists: SANN (\textbf{A}), RAD (\textbf{B}), Voronoi (\textbf{C}) and radial cut neighbor list (\textbf{D}). The points represent the values of PGOP for $\mathrm{C_2}$ and $\mathrm{C_3}$ point groups for pyrochlore crystal with Gaussian noise. The squares represent the values for the perfect crystal and ideal gas (black). The crystalline points and squares are colored according to their Wyckoff site designation. A perfect pyrochlore crystal is shown in \textbf{E}. Particles are colored according to their Wyckoff site, with their corresponding colors matching across the plots.}
    \label{fig:pyrochlore}
\end{figure*}

Next, we examined how the choice of neighbor list affects PGOP performance by computing PGOP values for a pyrochlore crystal (five Wyckoff sites) using four neighbor list methods from \textit{freud}: Solid-Angle Nearest Neighbors (SANN),\cite{van_meel_parameter-free_2012} Radial Angular Distance (RAD),\cite{higham_locally_2016,higham_overcoming_2018} Voronoi,\cite{voronoi_nouvelles_1908} and radial cutoff (1.1).

With the SANN neighbor list (Figure \ref{fig:pyrochlore} \textbf{A}), PGOP separates Wyckoff sites reasonably well in the perfect crystal. However, once noise is introduced, environments with the same PG symmetry (such as 16c and 16d ($\mathrm{S_6}$)) become indistinguishable. This is expected, as both sites share $\mathrm{C_3}$ as a subgroup, yielding identical PGOP values for $\mathrm{C_3}$ even without noise.

Surprisingly, the RAD neighbor list (Figure \ref{fig:pyrochlore} \textbf{B}) performs the worst, despite strong results in other systems shown so far. It assigns nearly identical PGOP values to multiple sites (e.g., 16c, 16d, and 32e) even when their point groups are unrelated. For instance, PGOP($\mathrm{C_2}$) is near 1 for both $\mathrm{C_3}$ and $\mathrm{S_6}$ sites, and PGOP($\mathrm{C_3}$) is also close to 1 for the low-symmetry 96g ($\mathrm{C_1}$). The large spread in values across environments in the noisy crystal renders RAD an ineffective choice when computing PGOP for pyrochlore.

Voronoi (Figure \ref{fig:pyrochlore} \textbf{C}) shows results similar to SANN: 16c and 16d are inseparable, but other environments—like 48f and 96g—remain distinguishable even under noise.

The radial cutoff list (RCut, Figure \ref{fig:pyrochlore} \textbf{D}) performs best overall in case of pyrochlore, enabling clear separation of all environments in both perfect and noisy systems, except for 16c and 16d environments (which share the same PG symmetry). These results demonstrate that with an appropriate neighbor list, PGOP can robustly distinguish local symmetries even in complex crystals under noisy conditions.

\subsection{Case study: Crystal nucleation in Lennard-Jones liquid}

\begin{figure*}
    \centering
    \includegraphics[width=\textwidth]{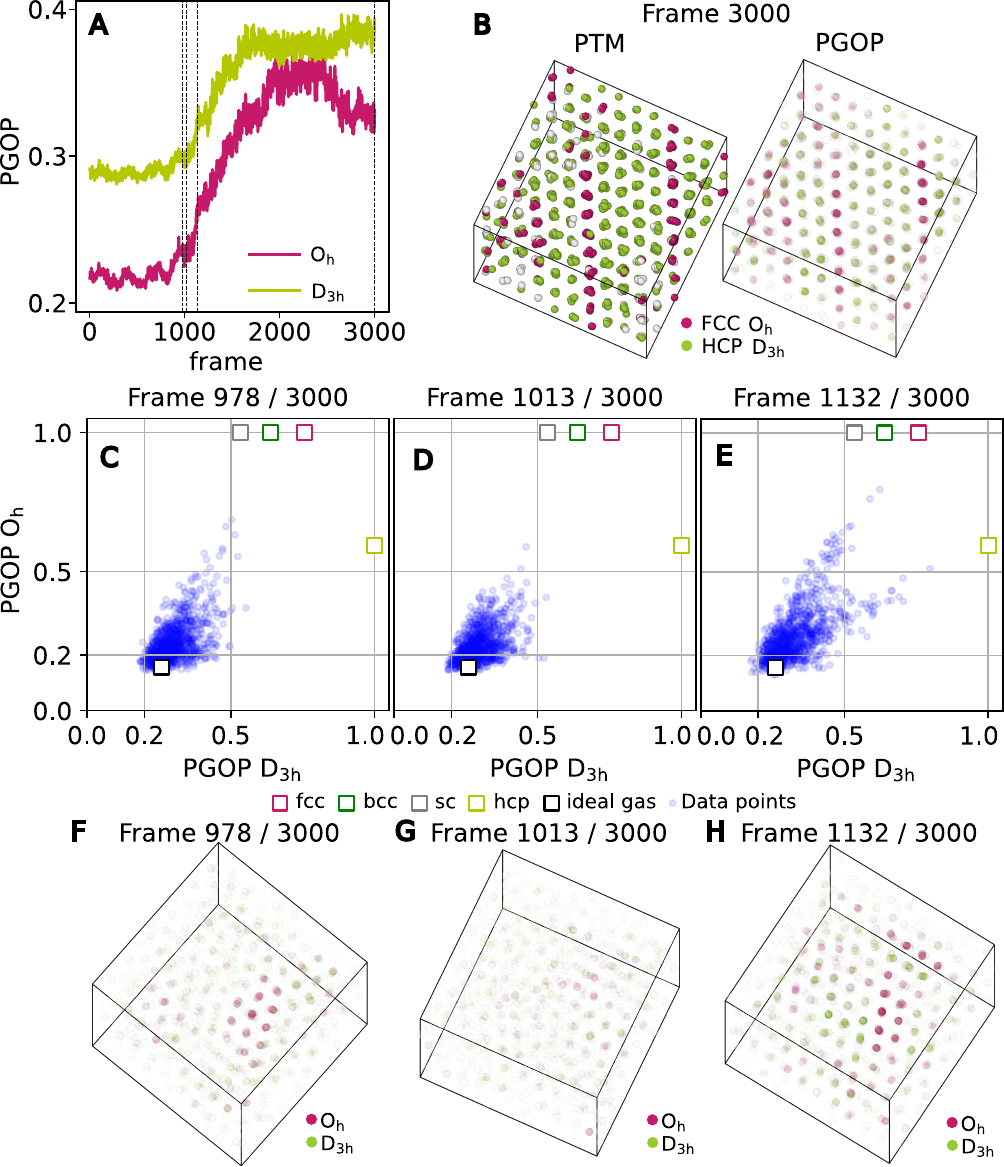}
    \caption{Nucleation of a crystal from a LJ liquid. System-averaged PGOP as a function of frame for the burst subtrajectory containing just the nucleation and crystallization frames is given in \textbf{A}. \textbf{B} shows the environment classification from Polyhedral Template Matching (left) and PGOP (right) for the final frame of the burst subtrajectory (marked with final dashed vertical line in \textbf{A}). The middle row contains the particle values of PGOP for point groups $\mathrm{D_{3h}}$ and $\mathrm{O_h}$ (labeled ``Data Points'') for following nucleation frames: frame 978 (\textbf{C} attempted unsuccessful nucleation), frame 1,013 (\textbf{D}  nucleus dissolution), and frame 1,132 (\textbf{E} successful nucleation).  PGOP values for perfect SC, BCC, FCC, HCP crystals and ideal gas are marked with a square with their respective colors. Location of these frames on the PGOP average plot are denoted with a dashed vertical lines. The final row contains the snapshots for these frames (\textbf{F}, \textbf{G} and \textbf{H} respectively). The particles are colored based on a $\frac{\mathrm{PGOP(D_{3h})}}{\mathrm{PGOP(O_h)}}$ ratio, with higher values corresponding to green for HCP ($\mathrm{D_{3h}}$) environments and lower values corresponding to magenta for FCC ($\mathrm{O_h}$) environments. Transparency used is given by the following formula: $\mathrm{Transparency}=1-(\mathrm{PGOP(D_{3h})}^2+\mathrm{PGOP(O_{h})}^2)/1.6$.}
    \label{fig:LJ}
\end{figure*}

Finally, we study the nucleation mechanism in a system of Lennard-Jones particles with molecular dynamics (MD). Crystallization of LJ particles is one of the best studied examples of crystallization.\cite{bulutoglu_comprehensive_2023,baidakov_spontaneous_2019,ouyang_entire_2020,rein_ten_wolde_numerical_1996,lundrigan_test_2009,baidakov_crystal_2010,trudu_freezing_2006}
We generate very high time-resolved (burst) subtrajectories of the nucleation event using the rare event detection library \textit{dupin}\cite{butler_change_2024} in conjunction with HOOMD-blue.\cite{anderson_hoomd-blue_2020}

We compute PGOP values for point groups $\mathrm{D_{3h}}$ and $\mathrm{O_{h}}$ for every particle in the system for the burst subtrajectory (Figure \ref{fig:LJ} \textbf{A}). The progression of the system-averaged values of PGOP shows that nucleation starts roughly around frame 1,000, where several key events are noted. First, in frame 978 (Figure \ref{fig:LJ} \textbf{F}), a small FCC nucleus is formed, characterized by several dozen data points of relatively high $\mathrm{PGOP(O_h)}>0.5$ (Figure \ref{fig:LJ} \textbf{C}). However, this nucleus dissolves quickly (frame 1,013, Figure \ref{fig:LJ} \textbf{G}) as evidenced by no particles with $\mathrm{PGOP}>0.5$ (Figure \ref{fig:LJ} \textbf{D}). Following this, a new nucleus is formed (frame 1,132, Figure \ref{fig:LJ}, \textbf{H}) in which we observe a core made of FCC-like particles with high $\mathrm{PGOP(O_h)}>0.5$ values, which successfully nucleate. This critical nucleus is also surrounded by HCP-like particles with high $\mathrm{PGOP(D_{3h})}>0.5$ (Figure \ref{fig:LJ} \textbf{E}), in contrast to the previous non-critical nucleus at frame 978. Based on this analysis, we conclude that under simulated conditions, a successful nucleation event begins with the appearance of HCP particles surrounding a FCC nucleus.\cite{bulutoglu_comprehensive_2023}

During the growth stage, both FCC and HCP environments grow together, showing classic HCP-FCC stacking faults. The process ends in an FCC crystal formed around frame 2,300. This is followed by a sizable amount of FCC particles transforming into HCP particles, resulting in a mostly HCP crystal system. We confirm this by visualizing (using OVITO\cite{stukowski_visualization_2010}) the last subtrajectory frame, frame 2,999. As seen in the top middle of Figure \ref{fig:LJ}, PTM shows an abundance (above $60\%$) of HCP particles in the system at this point. The PGOP values calculated for this same time step agree well with PTM based on the $\mathrm{PGOP(O_h)}$ and $\mathrm{PGOP(D_{3h})}$ values (Figure \ref{fig:LJ} \textbf{B}).

\section{Discussion}

\subsection{How to choose the distribution width and neighbor list for PGOP}

PGOP depends on two key parameters: the neighbor list and the distribution width ($\sigma$ for PGOP and $\kappa$ for PGOP-BOOD). These should be picked with care for the system studied.

The selection of $\sigma$ for PGOP and $\kappa$ for PGOP-BOOD can be justified by considering the intrinsic length scales of the system, such as the position of the first peak in the radial distribution function (RDF), as well as the degree of noise present, which is often correlated with temperature.
The distribution width should be chosen such that the distributions placed on the closest neighboring particles overlap minimally.
This approach ensures that the contribution of particle symmetrized to a closest neighbor particle gives negligible contributions for a point under consideration.
While narrower distributions are generally more sensitive to noise, they also impose higher computational costs.
Moreover, if the distribution widths become too small, the current optimization methodology may fail.
Therefore, parameter selection must strike a balance between sensitivity to structural variations and computational efficiency.
Users are recommended to choose the smallest feasible $\sigma$ (or largest $\kappa$) that maintains manageable computational costs and ensures successful optimization.

The neighbor list is another crucial factor influencing PGOP results.
While parameter-free neighbor lists such as SANN, RAD or Voronoi may seem appealing, they come with inherent tradeoffs.
Such lists, relying on geometric arguments rather than fixed cutoffs, can lead to inconsistencies in neighbor selection due to noise-induced positional fluctuations.
This variability often results in significantly broader OP distributions, which may not accurately reflect the noise in the underlying physical system.
Fixed-cutoff based neighbor lists address some of these inconsistencies by explicitly including particles within a predefined distance.
However, they are not immune to noise-related issues, particularly for systems where the first two peaks of the RDF overlap.
In such cases, both fixed cutoff and parameter-free methods can exhibit similar challenges in neighbor determination.

Considering PG symmetry order can be an effective way to understand the structure of particle neighborhoods in disordered phases as well.
Disordered systems may exhibit local PG symmetry at short distances that diminishes over longer distances, while crystals maintain long-range PG symmetry order.
To investigate these properties, the radial cutoff approach must be used, with similar considerations as previously discussed.
PGOP can also be applied to study PG symmetry order at varying length scales by adjusting the radial cutoff of the neighbor list. 
Parameter-free neighbor lists are generally better suited for analyzing short-range symmetry order due to their sensitivity to local structural features.

\subsection{Impact of distribution width on optimization}

The distribution width, $\sigma$, and likewise $\kappa$, also influences the optimization procedure.
Higher values of $\sigma$, or lower values of $\kappa$, require less computationally demanding optimization.
In practice we can get away with using fewer grid points for starting the brute-force search and still converge to the correct solution when $\sigma$ is large.
The optimal value of $\sigma$ should provide a satisfactory difference between the PGOP values of different local environments while maximizing its value for faster optimization.
For the simple crystals in Figure \ref{fig:simple_crystals}, we see that $\sigma=0.075$ provides a good balance between these two requirements.
A good rule of thumb is to choose the value of $\sigma$ for PGOP (or $\kappa$ value for PGOP-BOOD) such that the overlap between two distributions at the closest possible neighbor distance in the system should be very small.
Note that this choice may heavily depend on the relevant length scales in the system studied, and should be carefully chosen for each particular system.

\subsection{Modifications to PGOP required for differentiability with respect to particle coordinates}

An analytically differentiable OP with respect to particle positions could be very useful as a collective variable (CV) in advanced sampling methods such as metadynamics.\cite{laio_escaping_2002}
PGOP is already close to achieving full analytical differentiation (setting aside complications related to the differentiability of the neighbor-list choice), and with simple modifications it could be made fully differentiable with respect to particle coordinates.

The discrete nature of neighborhoods defined by the currently employed neighbor lists results in PGOP currently being non-continuous with respect to particle coordinates. To address this, a continuous formulation of PGOP can be achieved by removing strict distance cutoffs and implementing smooth transitions for particles entering and exiting the neighbor list. This approach can leverage weighted schemes for OP contributions, Voronoi neighbor lists, and face-area-based weighting for neighbor pairs.\cite{mickel_shortcomings_2013} Notably, this adjustment is feasible within the existing PGOP framework, as its code base already supports weighted neighbor lists.

The current method of determining the maximum overlap between symmetrized particle positions and all initial particle positions poses challenges for differentiability. A modification to use a summation of averaging over all initial positions, rather than a maximum value, would ensure complete differentiability. However, this adjustment might introduce a secondary effect where the PGOP values can exceed 1. This issue can be mitigated by using sufficiently narrow distributions, ensuring the upper limit remains close to 1.

Finally, achieving full differentiability would necessitate the removal of the optimization step. In specific scenarios, such as employing PGOP as a CV for metadynamics, this modification would even be advantageous. For instance, it might introduce a bias toward certain crystal orientations, which could align with box directions and thus, serve as a beneficial feature.

\section{Conclusion}

This work introduces a generalizable framework for constructing symmetry-based OPs. Earlier efforts characterized global (system-averaged) symmetry by assessing the point group order of global Bond-Orientational Order Diagrams (BOODs)\cite{engel_point_2021}, or by constructing local metrics through convolution of per-particle BOODs with Fisher distributions.\cite{butler_development_2024} Those BOOD-based approaches rely on truncated spherical harmonic expansions and numerical integration of inner products, both of which increase computational overhead and can reduce numerical precision in the OP. Although they can quantify arbitrary point-group symmetry of the BOOD via explicit symmetrization, they discard radial information by construction.\cite{butler_development_2024,engel_point_2021}
Our proposed framework instead operates directly in $\mathbb{R}^3$, retains radial information, and avoids spherical-harmonic basis expansions and truncation, enabling a more efficient implementation and a full-3D quantification of local point-group symmetry. As a result, PGOP is able to deliver significantly richer data than non-symmetry OPs like MSM (a variant of Steinhardt) while maintaining competitive performance.

The capability to track local, rather than global, symmetry makes PGOP particularly powerful for studying symmetry-breaking processes. By quantifying local point group symmetry continuously over time, one can observe the emergence of global order during crystallization and related phase transitions. PGOPs provide more direct quantification of PG symmetry order compared to all conventional bond orientational order parameters, offering better interpretability and improved sensitivity to structural order.

By tuning the choice of studied neighborhoods, PGOP can quantify either short-range or long-range order. Short-range order is particularly relevant for studying early-stage crystallization and the structure of disordered states such as liquids, glasses, and gels. In such cases, PG symmetry is typically assessed within the first coordination shell, which can be interpreted in terms of coordination polyhedra associated with the relevant PG symmetry (see Supplementary Material SM4).
The other tunable parameter in PGOP is the width of the distribution function, which affects its sensitivity. While an optimal value exists, it depends on system-specific factors as well as the neighbor list construction and the optimization scheme. 

PGOPs could also serve as collective variables for enhanced sampling approaches in the future. In this scenario, the optimization step can be skipped. Proper orientation allows the system to be biased toward crystal structures compatible with the simulation box, reducing grain boundary formation. Although the current implementation of PGOP is not differentiable, it requires only minimal modifications to become so.

Finally, PGOP and PGOP-BOOD calculations for all non-infinite PGs are implemented in the package SPATULA (Symmetry Pattern Analysis Toolkit for Understanding Local Arrangements), written in parallelized C++ with a Python interface.
This open-source implementation is freely available under a permissive license (BSD-3) at \url{github.com/glotzerlab/spatula}, with comprehensive documentation hosted at \url{spatula.readthedocs.io}.

\section*{Supplementary Material}
The supplementary material provides: definitions and 3D Cartesian representations of point-group symmetry operations; closed-form derivations of Bhattacharyya coefficients for Gaussian and Fisher distributions; a high-temporal-resolution burst subtrajectory for the Lennard–Jones nucleation event identified with \textit{dupin}; and representative polyhedral coordination environments with associated point group symmetries.

\section*{Acknowledgments}
This research was supported by the National Science Foundation, Division of Materials Research under a Computational and Data-Enabled Science and Engineering (CDS\&E) Award \# DMR 2302470, and by a Vannevar Bush Faculty Fellowship sponsored by the Department of the Navy, Office of Naval Research under ONR award number N00014-22-1-2821. This computational work used Anvil at Purdue through allocation DMR 140129 from the Advanced Cyberinfrastructure Coordination Ecosystem: Services \& Support (ACCESS) program, which is supported by National Science Foundation grants \#2138259, \#2138286, \#2138307, \#2137603, and \#2138296. Computational resources and services were provided by Advanced Research Computing at the University of Michigan, Ann Arbor.
We thank Dr. Joshua Anderson, Dr. Sun-Ting Tsai and Prof. Michael Engel, for valuable discussions and insightful suggestions during the course of this work.
We also thank Dr. Brandon Butler for his early contributions, published in his PhD dissertation,\cite{butler_development_2024} which inspired this work.

\section*{Author Declarations}
\subsection*{Conflict of Interest}
The authors have no conflicts to disclose.

\subsection*{Ethics Approval}
Ethics approval not required.

\subsection*{Author Contributions}
D.F.: Conceptualization, Methodology, Software, Validation, Formal analysis, Investigation, Visualization, Writing—original draft and editing.  
M.R.W.R.: Methodology, Validation, Software, Visualization, Writing—original draft and editing.
J.B.: Methodology, Validation, Software, Formal analysis, Investigation, Visualization, Writing—editing.
S.C.G.: Supervision, Funding acquisition, Resources, Writing—review and editing.

\section*{Data Availability}
The data that support the findings of this study are openly available at Deep Blue Data at \url{https://doi.org/10.7302/nxfa-zp21}. The code implementing the methodology developed in this work is available at \url{https://github.com/glotzerlab/spatula} under the BSD 3-Clause license.

\bibliographystyle{aipnum4-2}
\bibliography{references}

\end{document}


\title{Supplementary material: Quantifying Local Point-Group-Symmetry Order in Complex Particle Systems}

\author{Domagoj Fijan} 
\affiliation{Department of Chemical Engineering, University of Michigan, Ann Arbor, MI, USA}
\author{Maria R. Ward Rashidi}
\affiliation{Department of Materials Science \& Engineering, University of Michigan, Ann Arbor, MI, USA}
\author{Jenna Bradley}
\affiliation{Department of Materials Science \& Engineering, University of Michigan, Ann Arbor, MI, USA}
\author{Sharon C. Glotzer}
\affiliation{Department of Chemical Engineering, University of Michigan, Ann Arbor, MI, USA}
\affiliation{Department of Materials Science \& Engineering, University of Michigan, Ann Arbor, MI, USA}
\affiliation{Biointerfaces Institute, University of Michigan, Ann Arbor, MI, USA}

\maketitle

\section{Symmetry}
\label{SI:symmetry}

Symmetry can be described in terms of a group of transformations that leave the object invariant. This group is called the symmetry group of the object (or space).
There are several types of symmetry operations that are often encountered in physics, such as rotations, reflections, and translations. These operations can be combined to form more complex symmetry operations.

\subsection{Symmetry operators used in point groups}

In \textit{point} groups, the symmetry is defined with respect to a specific \textit{point} in space.
There are two main types of symmetry operations associated with point groups: rotations and reflections.
Alongside these, inversions and rotoreflections are also encountered.
Inversion can be considered as a special case of reflection, while rotoreflections can be constructed using a combination of rotations and inversions, or rotation and reflection.

Rotations are defined by the axis of rotation and the angle of rotation in three orthogonal spatial dimensions.
Rotations can be defined in many different ways, each with distinct advantages and disadvantages.
In PGOP, we primarily use the Euler angle representation in the zyz convention in keeping with the relevant literature\cite{altmann_symmetries_1957}.
In Schoenflies notation, a rotation operation is written as $\hat{C}_{nz}$, where $n$ is the order of rotation and the second letter indicates the rotation axis.
Other axes besides $x$, $y$, or $z$ can be used.
If the operation is written without an explicit rotation axis, such as $\hat{C}_n$, it denotes the main rotation axis (aligned with the rotation axis of the highest rotation order), typically taken to be the $z$ axis.
The angle of rotation ($\theta$) can be computed from the order $n$ using the formula:$\theta = 2\pi/n = 360^\circ / n$.
Multiple consecutive rotations are often applied in point groups and are written using power notation: $\hat{C}_n^m$.
This notation means that the operation $\hat{C}_n$ is applied $m$ times in succession.

Reflections are defined by the plane of reflection.
Reflections are a type of symmetry operation that flips the object across the plane of reflection.
Reflections cannot be written as rotations, or combination of rotations in a general sense.
Reflections can be written as an inversion followed by a rotation of 180 degrees
\cite{altmann_symmetries_1957}:
\begin{align}
    \hat{\sigma}_{xy} &= \hat{i} \hat{C}_{2z} \\
    \hat{\sigma}_h &= \hat{i} \hat{C}_2
\end{align}
where $\hat{i}$ is the inversion operator and $\hat{C}_{2z}$ is the two-fold rotation around the $z$ axis.
The reflection plane is always perpendicular to the axis of rotation obtained by the above formula.

Inversion is a symmetry operation that flips the object across the center of inversion.
It can be shown that inversion can be obtained by an application of 3 orthogonal reflections \cite{engel_point_2021}:
\begin{equation}
    \hat{i} = \hat{\sigma}_{yz} \hat{\sigma}_{xz} \hat{\sigma}_{xy}
\end{equation}
Rotoreflections are a combination of rotations and reflections, sometimes called improper rotations.
They are a type of symmetry operation that combines rotation and reflection. 
Thus, by definition, we can write \cite{altmann_symmetries_1957}:
\begin{equation}
    \hat{S}_n = \hat{\sigma}_h {\hat{C}_n} = \hat{\sigma}_{xy} {\hat{C}_n}
\end{equation}
where $\hat{\sigma}_h=\hat{\sigma}_{xy}$ is the reflection operator perpendicular to the axis of rotation ($z$).
Some useful equivalency relations for rotoreflections and their powers used in PGOP code can be found in work by Drago \cite{drago_physical_1992}.

\subsection{Symmetry Operation Representations}

In the study of PG symmetries in materials systems, it is essential to represent symmetry operations in a mathematical form that allows for their application to particle coordinates.
When symmetry operators act on vector spaces we talk about their representation.
This is typically achieved through matrix representations of the symmetry operations.
For our purposes, 3D Cartesian space is used to represent particle coordinates, and thus we write group action and symmetry operations in Cartesian (3D) representation.
Below we provide an overview of the key symmetry operations and their matrix representations for 3D Cartesian representation.

The identity matrix $\mathbf{I}$ represents the identity operation $\hat{E}$, which leaves all points unchanged:
\begin{equation}
\hat{E} =\mathbf{I} = 
\begin{pmatrix}
1 & 0 & 0 \\
0 & 1 & 0 \\
0 & 0 & 1
\end{pmatrix}.
\end{equation}

In three-dimensional space, a rotation matrix $\hat{C}_n$ can be constructed for a rotation by an angle $\theta$ about an axis defined by a unit vector $\vec{u} = (u_x, u_y, u_z)$.
More computationally efficient representations exist, such as quaternions, but PGOP uses the matrix representation for its simplicity and compatibility with other symmetry operations.
The general form of the rotation matrix in Cartesian coordinates is given by:
\begin{align}
&\hat{C}_n(\theta=2\pi/n, \vec{u}) =  \notag \\
&\begin{pmatrix}
\cos\theta + u_x^2(1 - \cos\theta) & u_x u_y (1 - \cos\theta) - u_z \sin\theta & u_x u_z (1 - \cos\theta) + u_y \sin\theta \\
u_y u_x (1 - \cos\theta) + u_z \sin\theta & \cos\theta + u_y^2(1 - \cos\theta) & u_y u_z (1 - \cos\theta) - u_x \sin\theta \\
u_z u_x (1 - \cos\theta) - u_y \sin\theta & u_z u_y (1 - \cos\theta) + u_x \sin\theta & \cos\theta + u_z^2(1 - \cos\theta)
\end{pmatrix}.
\end{align}
We use the implementation provided by SciPy to compute the rotation matrix from angle-axis representation or Euler angles in zyz notation.\cite{virtanen_scipy_2020}

Inversion is a symmetry operation that maps each point to its opposite point with respect to the origin.
The inversion matrix $\hat{i}$ in Cartesian representation is simply:
\begin{equation}
\hat{i} = -\mathbf{I} = 
\begin{pmatrix}
-1 & 0 & 0 \\
0 & -1 & 0 \\
0 & 0 & -1
\end{pmatrix}.
\end{equation}

The reflection matrix $\vec{\sigma}$ for a reflection across a plane with a normal vector $\vec{n} = (n_x, n_y, n_z)$ is given by:
\begin{equation}
\vec{\sigma} = \hat{i} \hat{C}_2 (\theta=\pi, \vec{n}).
\end{equation}

Rotoreflections are combinations of rotations and reflections and can be represented by composition of reflection and rotation.
A rotoreflection matrix $\vec{S}_n$ for a rotation by an angle $\theta=2\pi/n$ about an axis $\vec{u}$ followed by a reflection across a plane $\hat{\sigma}$ perpendicular to $\vec{u}$ can be constructed as:
\begin{equation}
\vec{S}_n (\theta=2\pi/n, \vec{u}) = \hat{\sigma} (\vec{u}) \hat{C}_n(\theta=2\pi/n, \vec{u}),
\end{equation}
where $\hat{\sigma}$ is the reflection matrix across the plane perpendicular to the axis of rotation and $\hat{C}_n$ is the rotation matrix.

\subsection{Symmetry point groups and their operations}

Infinitely many point groups exist.
Point groups are divided into categories according to the elements they contain and include the following:

\begin{itemize}
    \item Cyclic groups (starting with Schoenflies symbol C), which contain operations related to a rotation of a given degree $n$.
    \item Rotoreflection groups (S), which contain rotoreflection operations.
    \item Dihedral groups (D), which contain operations related to rotation of a given degree $n$ and reflection across a plane perpendicular to the rotation axis.
    \item Cubic/polyhedral groups (O, T, I), which contain symmetry operations related to important polyhedra in 3D space.
\end{itemize}

We give an overview of important point groups for materials science and crystallography below, with some remarks on notation and nomenclature.

With $\hat{\sigma}_h$ we label the reflection that is perpendicular (orthogonal) to the principal symmetry axis.
We label $\hat{\sigma}_v$ as the reflection parallel to the principal symmetry axis.
There are multiple choices one can make with parallel reflection, such as in the $zx$ or $zy$ plane.
With $\hat{\sigma}_d$ we usually label reflections parallel to the principal axis that are not $zx$ or $zy$.

The group operations are taken from the literature.\cite{cotton_chemical_1990,flurry_symmetry_1980}
We follow the nomenclature found in \cite{ezra_symmetry_1982} and \cite{altmann_semidirect_1963}.
In addition to that, we adopt a nomenclature in which $\hat{\sigma}_h = \hat{\sigma}_{xy}$ is the only horizontal reflection plane, while $\hat{\sigma}_{v}$ can be any reflection plane containing the principal axis of symmetry in the $z$ direction.
Note that some sources (such as \cite{ezra_symmetry_1982}) use $\hat{\sigma}^{'}$ for such reflection planes.
The designation $\hat{\sigma}_d$ denotes a subset of reflections $\hat{\sigma}_{v}$ which also bisect the angle between the twofold axes perpendicular to the principal symmetry axis ($z$).
We opt to not use the designation $\hat{\sigma}_d$.
The elements of groups $\mathrm{S_n}$ for odd values of $n$ are also given in \cite{drago_physical_1992}.

\begin{table}[h!]
\centering
\begin{tabular}{|c|l|}
\hline
\textbf{Point Group} & \textbf{Symmetry Operations} \\ \hline
$\mathrm{C_1}$ & $\hat{E}$ \\ \hline
$\mathrm{C_s}$ & $\hat{E}$, $\hat{\sigma}_v$ \\ \hline
$\mathrm{C_h}$ & $\hat{E}$, $\hat{\sigma}_h$ \\ \hline
$\mathrm{C_i}$ & $\hat{E}$, $\hat{i}$ \\ \hline
$\mathrm{C_n}$ & $\hat{E}$, $\hat{C}_n$, ${\hat{C}_n}^2$, ... ${\hat{C}_n}^{n-1}$ \\ \hline
$\mathrm{C_{nh}}$, $n$ is even & $\hat{E}$, $\hat{C}_n$, ${\hat{C}_n}^2$, ... ${\hat{C}_n}^{n-1}$, $\hat{\sigma}_h$, $\hat{S}_n$, ${\hat{S}_n}^3$, ... ${\hat{S}_n}^{n-1}$ \\ \hline
$\mathrm{C_{nh}}$, $n$ is odd & $\hat{E}$, $\hat{C}_n$, ${\hat{C}_n}^2$, ... ${\hat{C}_n}^{n-1}$, $\hat{\sigma}_h$, $\hat{S}_n$, ${\hat{S}_n}^3$, ... ${\hat{S}_n}^{2n-1}$ \\ \hline
$\mathrm{C_{nv}}$ & $\hat{E}$, $\hat{C}_n$, ${\hat{C}_n}^2$, ... ${\hat{C}_n}^{n-1}$, $n \hat{\sigma}_v$ \\ \hline
$\mathrm{D_n}$ & $\hat{E}$, $\hat{C}_n$, ${\hat{C}_n}^2$, ... ${\hat{C}_n}^{n-1}$, $n \hat{C}_2^{'}$ \\ \hline
$\mathrm{D_{nh}}$ & $\hat{E}$, $\hat{C}_n$, ${\hat{C}_n}^2$, ... ${\hat{C}_n}^{n-1}$, $n \hat{C}_2^{'}$, $\hat{\sigma}_h$, $\hat{C}_n \hat{\sigma}_h$, ${\hat{C}_n}^2 \hat{\sigma}_h$, ... ${\hat{C}_n}^{n-1} \hat{\sigma}_h$, $n\hat{\sigma}_v$ \\ \hline
$\mathrm{D_{nd}}$ & $\hat{E}$, $\hat{C}_n$, ${\hat{C}_n}^2$, ... ${\hat{C}_n}^{n-1}$, $n \hat{C}_2^{'}$, $\hat{S}_{2n}$, ${\hat{S}_{2n}}^3$, ... ${\hat{S}_{2n}}^{2n-1}$, $n\hat{\sigma}_v$ \\ \hline
$\mathrm{S_{n}}$, $n$ is even & $\hat{E}$, $\hat{S}_{n}$, ${\hat{S}_{n}}^2$, ... ${\hat{S}_{n}}^{n-1}$ \\ \hline
$\mathrm{S_{n}}$, $n$ is odd & $\hat{E}$, $\hat{S}_{n}$, ${\hat{S}_{n}}^2$, ... ${\hat{S}_{n}}^{2n-1}$ \\ \hline
$\mathrm{T}$ & $\hat{E}$, $4 \hat{C}_3$, $4 {\hat{C}_3}^2$, $3 \hat{C}_2$ \\ \hline
$\mathrm{T_h}$ & $\hat{E}$, $4 \hat{C}_3$, $4 {\hat{C}_3}^2$, $3\hat{C}_2$, $\hat{i}$, $3 \hat{\sigma}_h$, $4 \hat{S}_6$, $4 {\hat{S}_6}^5$ \\ \hline
$\mathrm{T_d}$ & $\hat{E}$, $8 \hat{C}_3$, $3 \hat{C}_2$, $6 \hat{\sigma}_v$, $6\hat{S}_4$ \\ \hline
$\mathrm{O}$ & $\hat{E}$, $6 \hat{C}_4$, $8 \hat{C}_3$, $9 \hat{C}_2$ \\ \hline
$\mathrm{O_h}$ & $\hat{E}$, $6 \hat{C}_4$, $8 \hat{C}_3$, $9 \hat{C}_2$, $3 \hat{\sigma}_h$, $6\hat{\sigma}_v$, $\hat{i}$, $8\hat{S}_6$, $6\hat{S}_4$ \\ \hline
$\mathrm{I}$ & $\hat{E}$, $12 \hat{C}_5$, $12 {\hat{C}_5}^2$, $20\hat{C}_3$, $15 \hat{C}_2$ \\ \hline
$\mathrm{I_h}$ & $\hat{E}$, $12 \hat{C}_5$, $12 {\hat{C}_5}^2$, $20\hat{C}_3$, $15 \hat{C}_2$, $15\hat{\sigma}_v$, $\hat{i}$, $12\hat{S}_{10}$, $12{\hat{S}_{10}}^3$, $20\hat{S}_6$ \\ \hline
\end{tabular}
\caption{Symmetry operations for various point groups}
\label{tab:symmetry_operations}
\end{table}

Notes on the table:

\begin{itemize}
    \item $\mathrm{C_{nv}}$: each $\hat{\sigma}_v$ is a reflection plane containing the principal axis of symmetry starting with $\hat{\sigma}_{yz}$, and the rest are successive rotations of the plane around $z$ axis by $\frac{\pi}{n}$.
    \item All dihedral groups ($\mathrm{D_n}$, $\mathrm{D_{nh}}$, $\mathrm{D_{nd}}$): each $\hat{C}_2^{'}$ is perpendicular to the principal axis of symmetry starting with $\hat{C}_{2x}$ and rest are successive rotation of this plane by $\frac{2\pi}{n}$.
    \item $\mathrm{D_{nh}}$: each $\hat{\sigma}_v$ is a reflection plane parallel to both principal ($z$) and each $\hat{C}_2^{'}$ axis.
    \item $\mathrm{D_{nd}}$: each $\hat{\sigma}_d$ is a reflection plane parallel to the principal axis of symmetry ($z$) and also contains the vector which bisects two neighboring $\hat{C}_2^{'}$ axes of symmetry.
    \item All tetrahedral groups ($\mathrm{T}$, $\mathrm{T_h}$, $\mathrm{T_d}$): see \cite{altmann_symmetries_1957} for specific proper rotations and also see Hurwitz quaternions.\cite{hurwitz_vorlesungen_1919}
    \item All octahedral groups ($\mathrm{O}$, $\mathrm{O_h}$): see Lipshitz\cite{dickson_theory_1924,Lipschitz1886} and Hurwitz\cite{hurwitz_vorlesungen_1919} quaternions for specific proper rotations.
    \item All icosahedral groups ($\mathrm{I}$, $\mathrm{I_h}$): see Hurwitz\cite{hurwitz_vorlesungen_1919} and icosian\cite{hamilton_memorandum_1856} quaternions for specific proper rotations.
\end{itemize}

In PGOP, all point groups are constructed from their operations given in the above table.

\section{Bhattacharyya coefficient derivation}
\label{SI:BCs}

The Bhattacharyya coefficient is a measure of the amount of overlap between two distributions.\cite{bhattacharyya_measure_1946,bhattacharyya_measure_1943}

\subsection{Bhattacharyya coefficient between two Gaussian distributions}

A normalized 3D Gaussian function is given by: 
\begin{equation}
G(\vec{r}, \sigma) =  \frac{1}{(2\sigma^2\pi)^{3/2}} \exp{-\frac{\lvert\vec{R} - \vec{r}\rvert^2}{2\sigma^2}}
\end{equation}
We use the Bhattacharyya coefficient (BC) to quantify the overlap between two Gaussians.
BC is defined as the integral of the square root of product of the two Gaussians:
\begin{equation}
BC(G_1,G_2) = \int_{\mathbb{R}^3} \sqrt{G_1G_2} dR
\end{equation}
The general formula for BC between two multivariate Gaussians is given by \cite{kashyap_perfect_2019}:
\begin{equation}
\mathrm{BC}(G_1,G_2)= \sqrt{\frac{\sqrt{\det{\Sigma_1}\det{\Sigma_2}}}{\det{\Sigma}}} \exp{-\frac{1}{8}(\vec{r}_1-\vec{r}_2)^T(\Sigma^{-1})(\vec{r}_1-\vec{r}_2)}
\end{equation}
where $\Sigma = \frac{1}{2}(\Sigma_1+\Sigma_2)$ and $\Sigma_1$ and $\Sigma_2$ are covariance matrices.
In our case the covariance matrices are diagonal, with elements $\sigma_1^2$ and $\sigma_2^2$ respectively.
This is because the width of our Gaussians is the same in every direction.
The determinants of the covariance matrices are given by $\det{\Sigma_1} = \sigma_1^6$, $\det{\Sigma_2} = \sigma_2^6$ and $\det{\Sigma}=\left(\frac{\sigma_1^2 + \sigma_2^2}{2}\right)^3$. Thus, the BC simplifies to:
\begin{equation}
\mathrm{BC}(G_1,G_2)= \left(\frac{2\sigma_1\sigma_2}{\sqrt{\sigma_1^2+\sigma_2^2}}\right)^{3/2} \exp{-\frac{\left|\vec{r}_1-\vec{r}_2\right|^2}{4\left(\sigma_1^2+\sigma_2^2\right)}}
\end{equation}
When two Gaussians are centered at the same point with identical standard deviation, the BC is 1.
If the two Gaussians are well separated with small standard deviations, the BC is close to zero.

\subsection{Bhattacharyya coefficient between two Fisher distributions}
In cases where only symmetries associated with local bond orientational order is of interest, a Fisher distribution (Gaussian on a sphere)\cite{fisher_dispersion_1953,watson_distributions_1982} can be employed.
The Fisher distribution is given by:
\begin{equation}
    P(\vec{r}, \kappa) = \frac{\kappa}{4\pi \sinh{\kappa}} \exp{\kappa \vec{r} \times \vec{R}}
\end{equation}
We can also derive the BC between two normalized Fisher distributions:
\begin{align}
   \mathrm{BC}(P_1,P_2) &= \int_{\mathbb{S}} \sqrt{P_1P_2} dS\\
    &= \sqrt{\frac{\kappa_1\kappa_2}{16\pi^2 \sinh{\kappa_1}\sinh{\kappa_2}}} \int_{\mathbb{S}} \exp{\left(\frac{1}{2}\left(\kappa_1 \vec{r}_1 \times \vec{R} + \kappa_2 \vec{r}_2 \times \vec{R}\right)\right)} dS \\
    &= \sqrt{\frac{\kappa_1\kappa_2}{16\pi^2 \sinh{\kappa_1}\sinh{\kappa_2}}} \int_{\mathbb{S}} \exp{\left(\frac{1}{2}\left(\kappa_1 \vec{r}_1 +\kappa_2 \vec{r}_2\right) \times \vec{R}\right)} dS
\end{align}
where we integrate over the unit sphere $\mathbb{S}$.
We can substitute $\vec{K} = \frac{\kappa_1\vec{r}_1 + \kappa_2\vec{r}_2}{2}$.
Notice that $\vec{K}$ is not normalized, even though $\vec{r}_1$, $\vec{r}_2$ and $\vec{R}$ are, because we are on the unit sphere.
The BC simplifies to:
\begin{equation}
    \mathrm{BC}(P_1,P_2) = \sqrt{\frac{\kappa_1\kappa_2}{16\pi^2 \sinh{\kappa_1}\sinh{\kappa_2}}} \int_{\mathbb{S}} \exp{\left(\vec{K} \times \vec{R}\right)} dS    
\end{equation}
The dot product of two vectors on a unit sphere is given by $\vec{K} \cdot \vec{R} = \left|\vec{K}\right| \cos{\theta}$, where $\theta$ is the angle between the two vectors.
Next, we convert to spherical coordinates by choosing $\vec{K}$ as the z-axis. We define $\theta$ as the polar angle and $\phi$ as the azimuthal angle, and the surface element on the sphere $dS = \sin{\theta} d\theta d\phi$:
\begin{equation}
    \mathrm{BC}(P_1,P_2) = \sqrt{\frac{\kappa_1\kappa_2}{16\pi^2 \sinh{\kappa_1}\sinh{\kappa_2}}} \int_{0}^{\pi} \int_{0}^{2\pi} \exp{\left(\left|\vec{K}\right| \cos{\theta}\right)} \sin{\theta} d\theta d\phi.
\end{equation}
The integral over $\phi$ is trivial and gives $2\pi$.
To evaluate the integral over $\theta$ we use the substitution $u = \cos{\theta}$, $du = -\sin{\theta} d\theta$:
\begin{equation}
    \mathrm{BC}(P_1,P_2) = -2\pi \sqrt{\frac{\kappa_1\kappa_2}{16\pi^2 \sinh{\kappa_1}\sinh{\kappa_2}}} \int_{1}^{-1} \exp{\left(\left|\vec{K}\right| u\right)} du.
\end{equation}
The integral is now a standard integral and its solution is $-\frac{2\sinh{|\vec{K}|}}{|\vec{K}|}$.
The final expression is then:
\begin{equation}
    \mathrm{BC}(P_1,P_2) = \sqrt{\frac{\kappa_1\kappa_2}{\sinh{\kappa_1}\sinh{\kappa_2}}} \frac{\sinh{|\vec{K}|}}{|\vec{K}|}.
\end{equation}
Finally, we can expand $|\vec{K}|$ to obtain the final expression, noting that $\vec{r}_1$ and $\vec{r}_2$ are normalized.
\begin{align}
    |\vec{K}| &= \frac{\sqrt{(\kappa_1\vec{r}_1 + \kappa_2\vec{r}_2)\times(\kappa_1\vec{r}_1 +\kappa_2\vec{r}_2)}}{2}\\
    &= \frac{\sqrt{\kappa_1^2\vec{r}_1\times\vec{r}_1 + \kappa_2^2\vec{r}_2\times\vec{r}_2 + 2\kappa_1\kappa_2\vec{r}_1\times\vec{r}_2}}{2}\\
    &=  \frac{\sqrt{\kappa_1^2+\kappa_2^2+2\kappa_1\kappa_2\vec{r}_1\times\vec{r}_2}}{2}.
\end{align}

Inserting this back into the BC expression we obtain:
\begin{equation}
    \mathrm{BC}(P_1,P_2) = 2 \sqrt{\frac{\kappa_1\kappa_2}{\sinh{\kappa_1}\sinh{\kappa_2}}} \frac{\sinh{\frac{\sqrt{\kappa_1^2+\kappa_2^2+2\kappa_1\kappa_2\vec{r}_1\times\vec{r}_2}}{2}}}{\sqrt{\kappa_1^2+\kappa_2^2+2\kappa_1\kappa_2\vec{r}_1\times\vec{r}_2}}.
\end{equation}
For identical Fisher distributions centered at the same position in space, the BC is 1, and for well separated Fisher distributions with high $\kappa$ the BC is close to zero.

\section{Burst trajectory of nucleation event using \textit{dupin}}

\begin{figure}
    \centering
    \includegraphics[width=\textwidth]{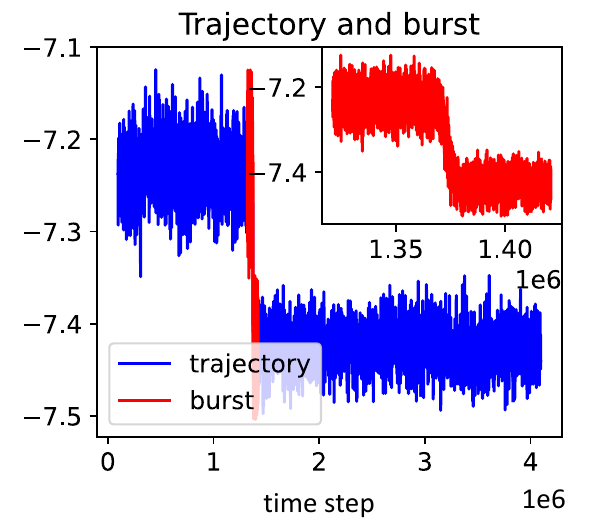}
    \caption{Subtrajectory of the rare nucleation event extracted using \textit{dupin}. The values plotted represent the 50th smallest potential energy values in the system at any given frame. The main plot shows the whole simulation trajectory with \textit{dupin} every 10000 steps, while the inset shows the zoomed-in burst subtrajectory that contains the nucleation event. Frames in this subtrajectory are spaced 10 time steps apart.}
    \label{sup:burst}
\end{figure}

Figure \ref{sup:burst} shows the obtained burst sub-trajectory of nucleation event with high temporal resolution using on-line change point detection with\textit{dupin}.

\section{Point group symmetries of some coordination polyhedra}
\label{SI:polyhedra_sym}
\begin{figure}
    \centering
    \includegraphics[width=\textwidth]{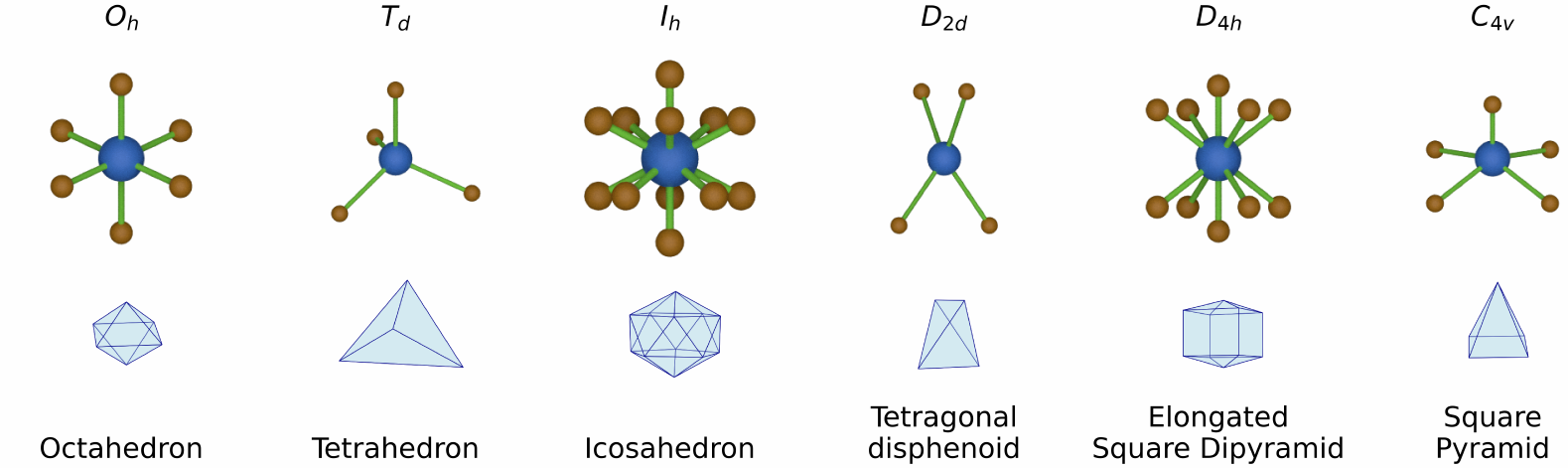}
    \caption{Examples of coordination environments corresponding to selected point groups. The top row displays ball-and-stick models illustrating coordination environments symmetric around the central blue particles. The bottom row shows the associated polyhedra, which share the same PG symmetries as their respective coordination environments.}
    \label{fig:coordination}
\end{figure}
Examples of key coordination polyhedra and their PG symmetries are given in Figure \ref{fig:coordination}. Note that interpreting PGOP results as symmetry of coordination polyhedra will be highly dependent on the neighbor list choice. This interpretation can be particularly helpful for point groups other than the polyhedral groups (T, $\mathrm{T_h}$, O, $\mathrm{O_h}$, I and $\mathrm{I_h}$), as these six groups can have many different regular polyhedra associated with them, which is reflected in their designations as polyhedral groups.

\section{Performance and Parallelism}
\label{SI:performance}

\begin{figure}
    \centering
    \includegraphics[width=\textwidth]{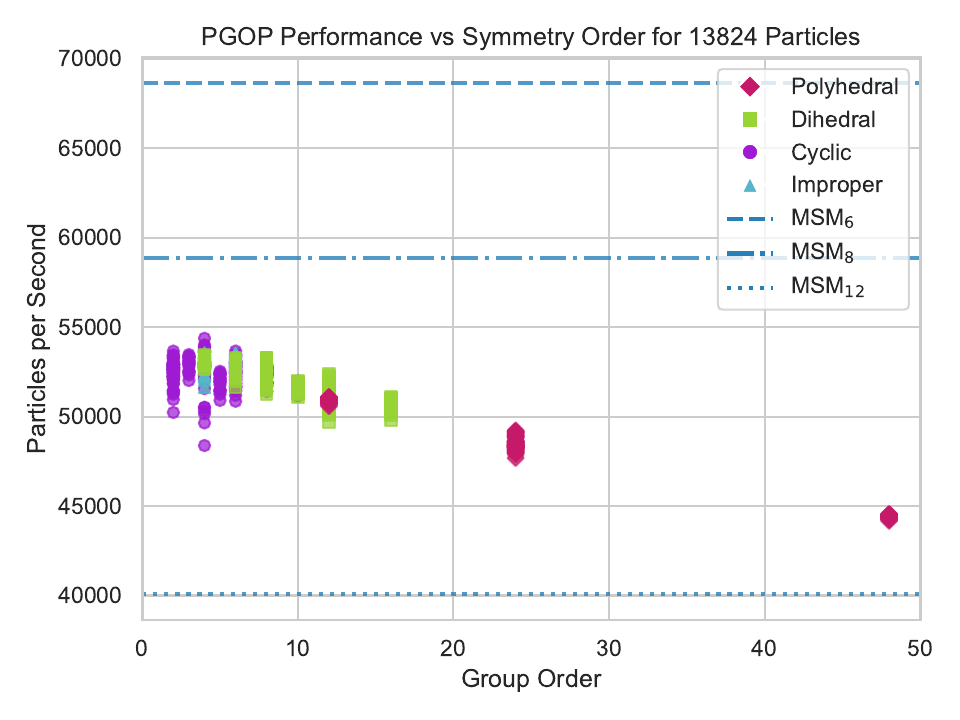}
    \caption{Performance of PGOP as a function of group order $G$, highlighting the performance similarity to several $\mathrm{MSM}_l$ order parameters. Measurements were taken on a single thread with no $SO(3)$ optimization.}
    \label{sup:performance_grouporder}
\end{figure}

The SPATULA library makes use of both thread-based and instruction-level parallelism to accelerate the calculation of PGOP and PGOP-BOOD order parameters. This enables superlinear scaling in the performance with respect to group size, as evidenced by Figure \ref{sup:performance_grouporder}, a direct result of the SIMD instructions we employ. This allows even large point groups like $\mathrm{O_h}$ to be evaluated in a similar time to $\mathrm{MSM}_12$ when no optimization is required, suggesting future improvements to the optimization procedure could yield even greater performance. We also note that all benchmarks performed for this paper made use of 128 or 256-bit wide SIMD registers, rather than the 512-bit wide instructions that our code currently supports. Users with access to this hardware feature should expect improved performance on both single-core and parallel jobs.

\begin{figure}
      \centering
      \includegraphics[width=\textwidth]{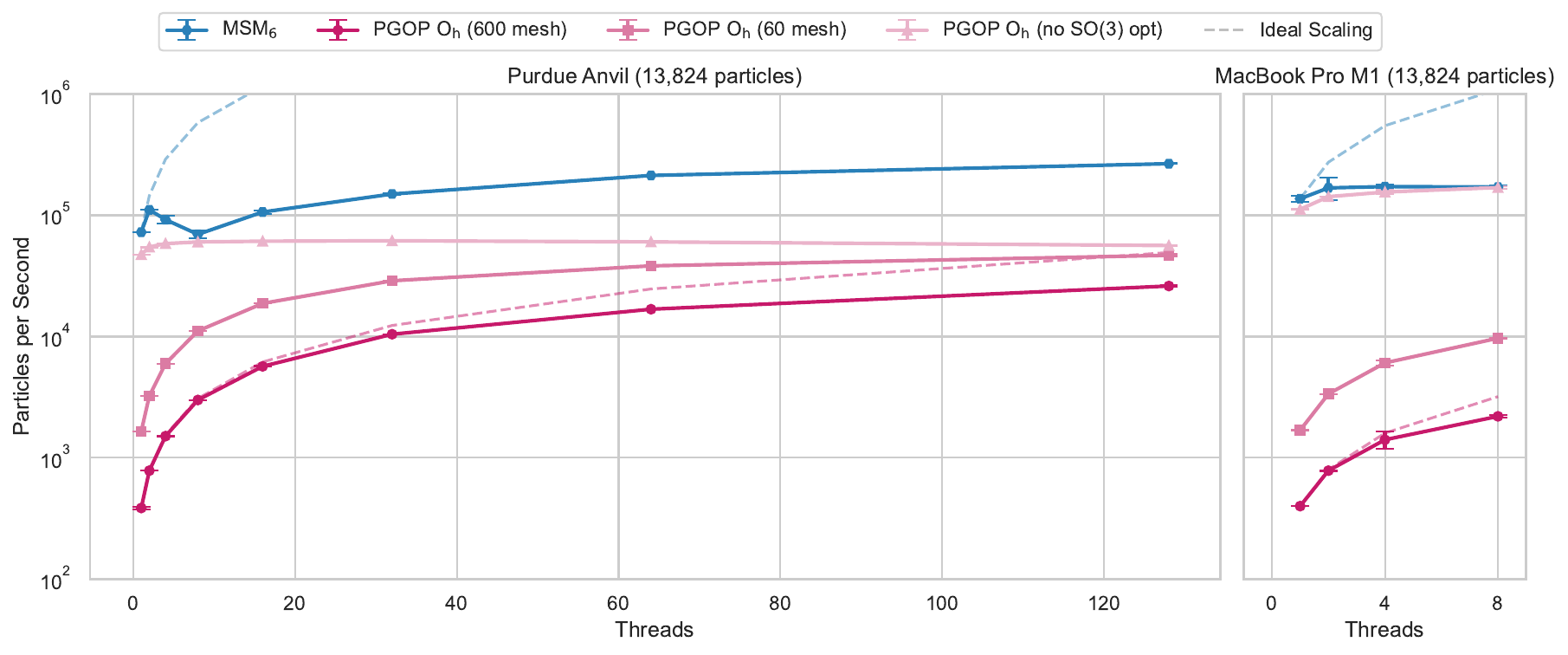}
      \caption{Strong scaling performance comparison of $\mathrm{MSM}_6$ with PGOP $\mathrm{O_h}$. Left: Purdue Anvil. Right: M1 Macbook. Dashed lines show  ideal parallel scaling.}
      \label{sup:performance}
  \end{figure}

Figure \ref{sup:performance} shows the parallel scaling of our implementation of the PGOP order parameter, on both the Purdue Anvil supercomputer as well as more modest consumer hardware. Excellent parallel scaling is observed up to about 32 cores, at which point the efficiency drops from $84.7\%$ to $68.1\%$. At this point, the quantity of work per benchmark iteration is no longer sufficient to saturate the entire thread pool. Larger systems display clear parallel scaling up to the full 128 cores.

\section{Sensitivity to thermally induced noise}
\label{sec:realistic_melt}

\begin{figure}
      \centering
      \includegraphics[width=\textwidth]{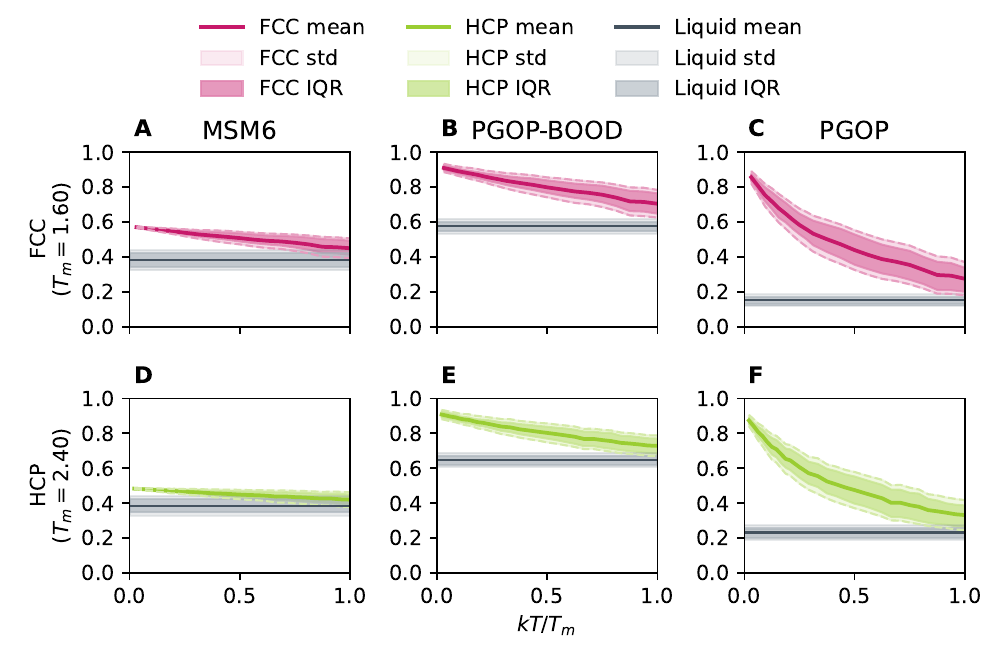}
      \caption{Order measured by MSM ($l=6$), PGOP, and PGOP-BOOD for FCC and HCP crystals with increasing thermal noise in LJ systems given in units relative to melting temperature ($T_m$). Line and shaded-region colors indicate FCC (magenta), HCP (yellow-green), and liquid at the melting temperature (black/gray). For each dataset, the darker shaded region shows the interquartile range (first to last quartile), and the lighter shaded region with the dashed line shows $\pm$ one standard deviation. Panels \textbf{A} and \textbf{D} show MSM$6$ for FCC and HCP, respectively. For FCC (top row), PGOP-BOOD (\textbf{B}) and PGOP (\textbf{C}) are computed for the point group $\mathrm{O_h}$. For HCP (bottom row), PGOP-BOOD (\textbf{E}) and PGOP (\textbf{F}) are computed for the point group $\mathrm{D{3h}}$.}
      \label{sup:melting_simple_crystals}
\end{figure}

The thermally induced noise results show a clear and systematic degradation of crystalline order with increasing temperature for both FCC and HCP LJ systems, while preserving strong discrimination among the three order metrics across most of the pre- melting regime ($kT/T_m <0.4$). All order parameter values decrease monotonically, but with distinct sensitivities: MSM$_6$ declines earlier and approaches the disordered liquid baseline at around 0.4 for HCP and 0.75 for FCC, whereas PGOP and PGOP-BOOD remain separated from the liquid reference all the way up until the melting temperature ($T_m$). This delayed collapse is visible both in the mean trends and in distribution-level statistics (IQR and standard-deviation envelopes), indicating that the improved discrimination is not an artifact of averaging. As expected, near ($kT/T_m!\approx!1$) all crystal-branch curves converge toward liquid values, consistent with loss of long-range crystalline signatures; however, the PGOP-family descriptors maintain larger margins to baseline over a wider thermal window than MSM$_6$, supporting the central claim that they provide more robust structural discrimination under strong thermal noise.

\bibliographystyle{unsrt}
\bibliography{references}